\documentclass[journal,comsoc]{IEEEtran}

\usepackage[T1]{fontenc}
\ifCLASSINFOpdf

\else

\fi

\usepackage{amsmath}
\usepackage{supertabular}
\usepackage{comment}
\usepackage{cite}
\usepackage{color}
\usepackage{graphicx}
\usepackage{epsfig,graphics,subfigure,psfrag,amsmath,amssymb}
\usepackage{stfloats}
\usepackage{amsfonts}
\usepackage{slashbox}
\usepackage{multirow}
\usepackage{amssymb}
\usepackage{algorithmic}
\usepackage{url}
\usepackage{hyperref}
\usepackage{breakurl}
\usepackage{bm}
\usepackage{mathrsfs}
\usepackage{amssymb}
\usepackage{makecell}
\usepackage{array}
\usepackage{textcomp}

\begin{document}

\title{IEEE 802.11ay based mmWave WLANs: Design Challenges and Solutions}

\author{Pei~Zhou, Kaijun~Cheng, Xiao~Han,
        Xuming~Fang,~\IEEEmembership{Senior Member,~IEEE,}
        Yuguang~Fang,~\IEEEmembership{Fellow,~IEEE,}
        Rong~He, Yan~Long,~\IEEEmembership{Member,~IEEE,} and Yanping~Liu

\thanks{P. Zhou, K. Cheng, X. Fang, R. He, Y. Long and Y. Liu are with Key Lab of Information Coding \& Transmission, Southwest Jiaotong University, Chengdu 610031, China (e-mails: peizhou@my.swjtu.edu.cn; kaijuncheng@my.swjtu.edu.cn; xmfang@swjtu.edu.cn; rhe@swjtu.edu.cn; yanlong@swjtu.edu.cn; liuyanping@my.swjtu.edu.cn). \emph{(X. Fang is the corresponding author.)}} 
\thanks{X. Han is with the CT lab, Huawei, Shenzhen 518129, China (e-mail: tony.hanxiao@huawei.com).}%
\thanks{Y. Fang is with the Department of Electrical and Computer Engineering, University of Florida, 435 New Engineering Building, PO Box 116130, Gainesville, FL 32611, USA. (e-mail: fang@ece.ufl.edu).}
}

\markboth{IEEE Communications Surveys \& Tutorials,~Vol.~XX, No.~XX, XXX~201X}
{}

\maketitle

\begin{abstract}
Millimeter-wave (mmWave) with large spectrum available is considered as the most promising frequency band for future wireless communications. The IEEE 802.11ad and IEEE 802.11ay operating on 60 GHz mmWave are the two most expected wireless local area network (WLAN) technologies for ultra-high-speed communications. For the IEEE 802.11ay standard still under development, there are plenty of proposals from companies and researchers who are involved with the IEEE 802.11ay task group. In this survey, we conduct a comprehensive review on the medium access control layer (MAC) related issues for the IEEE 802.11ay, some cross-layer between physical layer (PHY) and MAC technologies are also included. We start with MAC related technologies in the IEEE 802.11ad and discuss design challenges on mmWave communications, leading to some MAC related technologies for the IEEE 802.11ay. We then elaborate on important design issues for IEEE 802.11ay. Specifically, we review the channel bonding and aggregation for the IEEE 802.11ay, and point out the major differences between the two technologies. Then, we describe channel access and channel allocation in the IEEE 802.11ay, including spatial sharing and interference mitigation technologies. After that, we present an in-depth survey on beamforming training (BFT), beam tracking, single-user multiple-input-multiple-output (SU-MIMO) beamforming and multi-user multiple-input-multiple-output (MU-MIMO) beamforming. Finally, we discuss some open design issues and future research directions for mmWave WLANs. We hope that this paper provides a good introduction to this exciting research area for future wireless systems.
\end{abstract}


\IEEEpeerreviewmaketitle

\section{Introduction}
\IEEEPARstart{R}{ecently}, IEEE 802.11ad (or 802.11ad for short), a Wireless Fidelity (Wi-Fi) standard for wireless local area networks (WLAN) to provide data throughput rates of up to 6 gigabits per second (Gbps) at frequencies around 60 GHz, has rolled out as the newest member of the WLAN 802.11 family. However, similar to any other communications standards, due to emerging applications or service requirements, evolution always follows. A recent enhancement, dubbed \emph{IEEE 802.11ay} (\emph{802.11ay} for short), that promises to deliver faster and longer range wireless transmissions, has already under development. The highlight of this new standard is its very high performance for fixed point-to-point (P2P) and point-to-multipoint (P2MP) transmissions, either indoor or outdoor, to meet the requirements of people's daily information exchange and entertainments. The 802.11ay standard task group states its goal as follows: ``\emph{Task Group ay is expected to develop an amendment that defines standardized modifications to both the 802.11 physical layers (PHY) and the 802.11 medium access control layer (MAC) that enables at least one mode of operation capable of supporting a maximum throughput of at least 20 gigabits per second, while maintaining or improving the power efficiency per station (STA).}'' \cite{ref1}. It indicates that 802.11ay is an evolution instead of revolution of 802.11ad. Due to the page limit, this paper will mainly track the evolution and advancements from the perspective of MAC and MAC related cross-layer design issues between PHY and MAC. We are attempting to answer the following most important questions: what is the motivation developing 802.11ay? What are the key differences between 802.11ad and 802.11ay? What are the new technologies such as the enhanced channel aggregation and channel bonding, channel access and allocation, MIMO and beamforming? How do they work? What are the potential technologies to be developed and the remaining unfolding design challenges behind 802.11ay? For better understanding of the structure of this paper, the specific relationships among these technologies are shown in Fig. 1, which is based on 802.11ad standard and 802.11ay draft \cite{ref2, ref3, ref4}. We observe from Fig. 1 that the two important PHY related techniques, namely, channel bonding and channel aggregation, are included. For the MAC related techniques, we mainly introduce beamforming training (BFT) and channel access \& allocation. There are many new features in both BFT and channel access \& allocation. In addition, some open issues are also introduced to provide possible future research directions for our readers.

\begin{figure*}[!htbp]
  \begin{center}
    \scalebox{0.85}[0.85]{\includegraphics{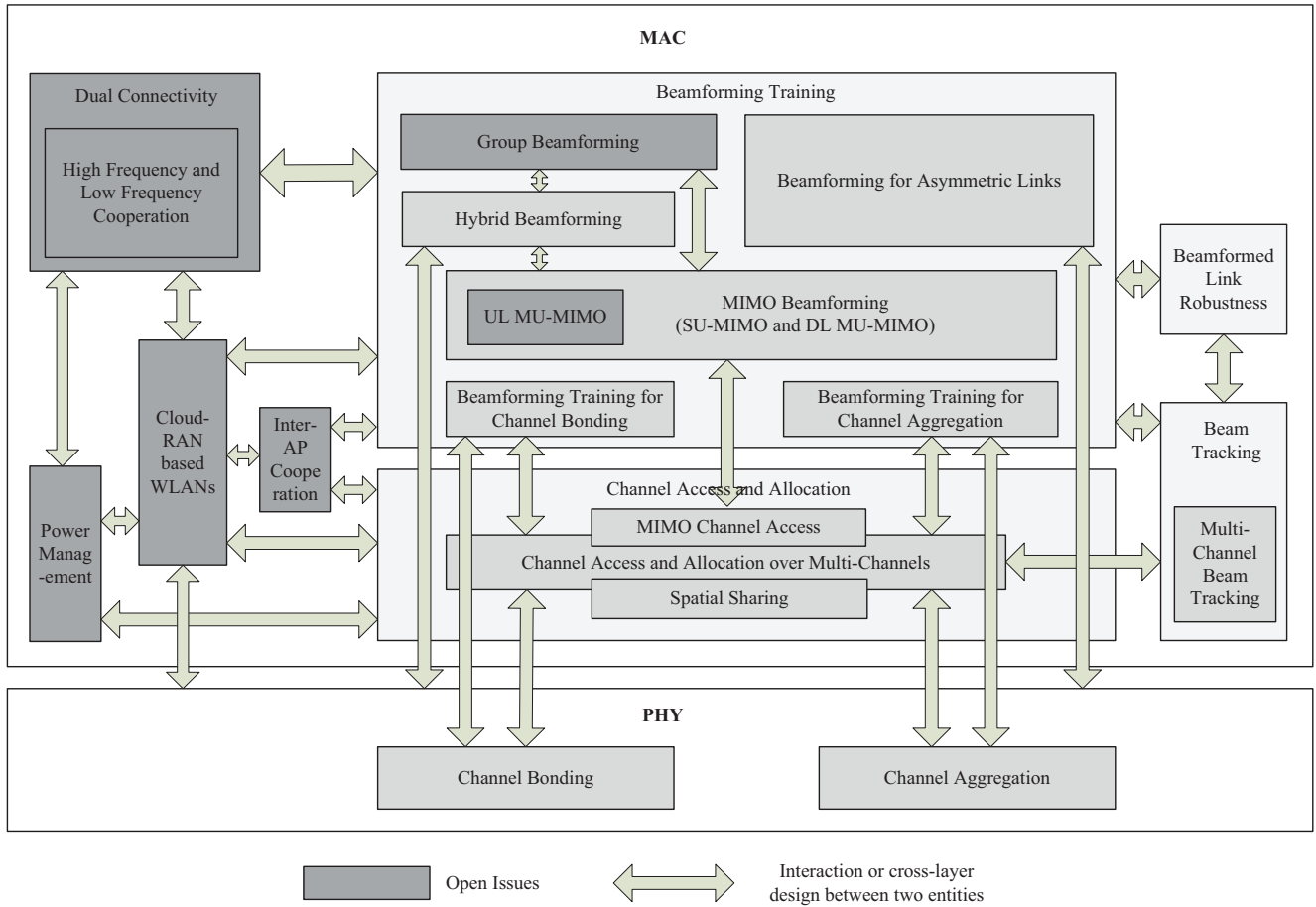}}
    \caption{Overview of the related technologies and their interactions in this survey.}
  \end{center}
\end{figure*}

Since there are limited unlicensed spectra in low frequency (LF) band (e.g., 2.4 GHz and 5 GHz, etc.), which has already been congested, the existing 802.11 WLANs (e.g., 802.11n, 802.11ac, etc.) could only offer relatively low data rate for new emerging wireless applications. In order to provide higher data rates for users to meet the ever-increasing high-speed wireless transmissions, more spectra and more advanced technologies are needed. Millimeter-wave (mmWave) (i.e., 30 GHz - 300 GHz) with tremendous available spectra \cite{ref5} can offer new opportunities for broadband applications. Currently, mmWave communication has been viewed as one of the most promising technologies for the fifth generation (5G) mobile communication systems \cite{ref6, ref7}, and some advanced short-range communication systems, such as the IEEE 802.15.3c \cite{ref8}, 802.11ad \cite{ref2, ref3} and 802.11ay \cite{ref4}, etc. Within these rich unlicensed spectra, the 60 GHz unlicensed mmWave was recommended as the most promising frequency band \cite{ref9}, in which there are four 2.16 GHz channels supported in 802.11ad. However, the 802.11ad is only permitted to choose one 2.16 GHz channel to use \cite{ref2, ref3}. That is to say, it does not support multi-channel operations, which limits the flexibility and efficiency of the use of four channels. To support higher throughput, transmissions through wider spectrum (e.g., a bonded channel or an aggregated channel) are considered for 802.11ay \cite{ref4}.

Since mmWave propagation suffers severe path-loss and signal attenuation, it is hardly possible to achieve long distance communications when adopting conventional omni-directional transmissions. Therefore, beamforming technology, which can concentrate the transmit power and the receive region over narrow beams, suitable for directional transmissions, becomes the core technology for mmWave communications \cite{ref10}. However, the BFT processes have to be carefully designed to determine the appropriate transmit and receive beam before directional transmissions. These processes are usually time-consuming. Thus, a low complexity but efficient BFT method will be of great importance, especially for WLANs which aim to provide high quality of services (QoS) with low cost and low complexity. Although the BFT processes in 802.11ad are already efficient to some extent, there are still many remaining design issues to tackle for 802.11ay. The 802.11ay task group was formed in May 2015 \cite{ref4, ref11}, and it has almost completed its tasks.

WLAN technologies have already received considerable attention in the past 20 years, and there are already many good surveys on 802.11 WLANs. Zhu \emph{et al.} in \cite{ref12} presented a survey on the QoS in 802.11 WLANs, including admission control, bandwidth reservation in MAC and higher layers, service differentiation in MAC, and link adaptation in PHY. Thorpe and Murphy discussed the adaptive carrier sensing mechanisms for 802.11 WLANs in \cite{ref13}, which provides a detailed review on the optimal physical carrier sensing, adaptive physical carrier sensing in 802.11 wireless ad hoc networks and infrastructure networks. However, these surveys are based on the traditional low frequency 802.11 WLANs, which are very different from the WLANs in high frequency (HF, e.g., 60 GHz mmWave). There is also a survey on PHY/MAC enhancements and QoS mechanisms for Very High Throughput WLANs \cite{ref14} under four different WLANs standards, namely 802.11n, 802.11ac, 802.11ad and 802.11aa. In \cite{ref15}, a comprehensive survey was conducted on the design issues by taking the 802.11ad into consideration, such as the PHY, MAC network architecture and beamforming protocols, etc. In addition, some novel technologies are also elaborated to overcome the challenges of mmWave communications. However, to the best of our knowledge, there is currently no comprehensive survey on the newly designed 802.11ay. Thus, in this paper, we will focus on the MAC related and some cross-layer technologies of 802.11ay, such as channel bonding and aggregation, channel access and allocation for multiple channels, efficient BFT and beam tracking, Single-User Multiple-Input-Multiple-Output (SU-MIMO) and Multi-User Multiple-Input-Multiple-Output (MU-MIMO) operations, etc., which are the most important technologies to improve the QoS and higher throughput.

The remainder of this paper is organized as follows: Section II introduces the relevant standards and technical design challenges. Channel aggregation and bonding for 802.11ay are discussed in Section III. Section IV provides channel access and allocation for 802.11ay. In Section V, the efficient BFT and beam tracking methods are presented. Section VI describes SU-MIMO and MU-MIMO beamforming. Section VII discusses some open issues and future research directions. Finally, we conclude the survey in Section VIII. Since the 802.11ay standard draft 1.0 has already been released and there were almost no major changes compared to the previous standard drafts on the technical aspects. Therefore, we think that the survey will not have the risk of becoming obsolete or technically inaccurate when the final standard is released in the future.

For better understanding, the terms and acronyms used throughout the whole paper are listed in Appendix A.

\section{Relavent Standards Evolution and Technical Challenges}
Since 802.11ay is an enhancement of the 802.11ad standard completed in 2012 \cite{ref16}, in this section, we firstly summarize the MAC related technologies of 802.11ad. Then, we elaborate some technical challenges on mmWave communications. Finally, we summarize some important features and technologies in the 802.11ay.

\subsection{MAC related issues in the 802.11ad standard}
Directional Multi-Gigabit (DMG) channel access and DMG BFT are the two most important techniques in the 802.11ad, we first briefly describe their procedures for the 802.11ad. More details can be found in \cite{ref2}.

\subsubsection{DMG channel access}

As described in \cite{ref2}, DMG channel access in 802.11ad is coordinated by the DMG access point (AP) or the personal basic service set (PBSS) control point (PCP) according to beacon interval (BI) timing according to an appropriately designed schedule. The scheduling information can be included in DMG Beacon and Announce frames. After receiving such scheduling information, DMG STAs access the medium based on the access rules according to the specific periods of BI. As shown in Fig. 2, a BI is generally comprised of a beacon header interval (BHI) and a data transfer interval (DTI). The BHI contains a beacon transmission interval (BTI), an association beamforming training (A-BFT), and an announcement transmission interval (ATI), while the DTI consists of contention-based access periods (CBAPs) and scheduled service periods (SPs). Note that any combination in the number and order of CBAPs and SPs can be present in the DTI.

\begin{figure}[!htbp]
  \begin{center}
    \scalebox{0.42}[0.42]{\includegraphics{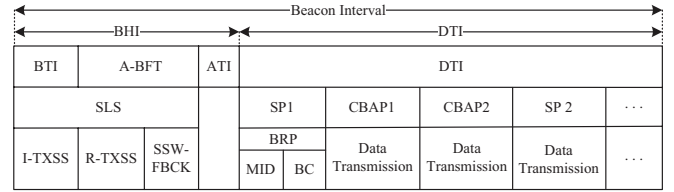}}
    \caption{Example of access periods within a BI.}
  \end{center}
\end{figure}

In the BTI, the DMG PCP/AP transmits the DMG Beacon frames to perform initiator transmit sector sweep (I-TXSS). In the A-BFT, DMG STAs perform responder transmit sector sweep (R-TXSS) by using sector sweep frames (SSW frames). During ATI, the PCP/AP exchanges management information with associated and beam-trained STAs, and only the PCP/AP can initiate a frame transmission. The frames transmitted in ATI are limited to the request and response frames such as Management frame, Ack frame, etc. The details of specific request and response frames can be found in \cite{ref2}. In particular, ATI is optional, and it is indicated by the ATI Present field that is set to 1 in the current DMG Beacon frame. DTI is mainly a data transmission period. It is noted that not all the DMG STAs have the opportunity to transmit during DTI. A DMG STA which intends to initiate a frame exchange in DTI shall meet one of the following two conditions:

(1) during a CBAP in which the STA is identified or included as source or destination,

(2) during an SP in which the STA is identified as source or destination.

In CBAP, multiple DMG STAs compete for the medium according to the 802.11 enhanced distributed coordination function (EDCF). An SP is allowed for communication between a dedicated pair of DMG STAs in a contention-free period \cite{ref15}. It is worth noting that, since 802.11ad does not support multi-channel operations, all the operations described above in DMG channel access are limited to a single channel (a primary channel).

\subsubsection{DMG beamforming}

In order to compensate the serious path loss in 60 GHz mmWave band and provide necessary DMG link budget for a pair of STAs, beamforming becomes an essential procedure since it can concentrate the transmit (TX) power and receive (RX) region on a relatively narrow beam. The two main components of BFT in 802.11ad are Sector-level sweep (SLS) phase and Beam Refinement Protocol (BRP) phase as shown in Fig. 2. SLS can take place in both `BTI + A-BFT' phase and DTI phase. In BTI, the DMG PCP/AP uses DMG Beacon frames and DMG STAs use sector sweep frames to train their TX sectors. However, BRP can only be performed in DTI phase and it trains the RX sectors of both initiator and responder. BRP can also improve both TX antenna configuration and RX antenna configuration through iterative procedures. It is composed of BRP setup subphase, Multiple sector ID Detection (MID) subphase, Beam Combining (BC) subphase, and one or more beam refinement transactions. The details of SLS and BRP are described in Section V and \cite{ref2}.

In the 802.11ad, DMG beamforming does not support MU BFT concurrently. SU-MIMO beamforming, MU-MIMO beamforming and hybrid beamforming are also not supported. Collision is a serious shortage in A-BFT which should be relieved in dense user scenarios. Fortunately, there are many improvements and enhancements on beamforming proposed in the 802.11ay proposals, standard draft and related literatures, which will be introduced in Sections V and VI.

\subsection{Technical challenges on mmWave communications}
Although in theory mmWave communications can provide ultra-high-speed data rate with large available spectrum, there are many practical design issues that need to be addressed seriously. In this subsection, we discuss some technical challenges on mmWave communications, including topics such as channel aggregation, bonding and allocation; BFT and beam tracking; beamformed link blockage; spatial sharing (SPSH) and interference mitigation; power management. In addition, more open design issues and challenges for future mmWave communication systems will be further elaborated in Section VII.

\subsubsection{Channel aggregation, bonding and allocation}

There is no doubt that channel bonding and channel aggregation are promising technologies to provide significant throughput gains with lower signal-to-interference-and-noise ratio (SINR) due to the reduction of the transmission power per hertz each time the channel width is doubled. However, it also brings in some serious problems. Deek \emph{et al.} \cite{ref17} and Arslan \emph{et al.} \cite{ref18} experimentally analyzed the advantages and disadvantages of channel bonding in 802.11n. They identified the following problems:

(1) a lower transmission range because wider channels require higher sensitivity,

(2) a higher probability to suffer from and result in interference,

(3) more competitions for channels with other WLANs operating in the same area.

In addition, since both the channel center frequency and the channel width are autonomously selected by each WLAN, the use of channel bonding will increase the probability of spectrum overlapping among WLANs operating in the same area. Spectrum overlapping means that several BSSs share at least one basic channel, which may lead to significant performance degradation for some or all of them.

In terms of channel bonding for mmWave, the up-to-GHz wideband characteristics of the millimeter wave frequency present new challenges except for the typical challenges of microwave. Thus, mmWave demands new features in channel bonding and channel access. We summarize some of the key design challenges related to channel bonding in mmWave frequency band as follows:

(1) New design at both MAC and PHY is indispensable. For example, the possible enhancements for channel coding scheme to support channel bonding and MIMO in orthogonal frequency division multiplexing (OFDM) PHY, which can provide packet error rate (PER) performance improvement.

(2) Need for new efficient BFT schemes for contiguous channel bonding and non-contiguous channel aggregation, particularly in combining with SU-MIMO and MU-MIMO scenarios.

(3) New operations of channel access with channel bonding and channel aggregation should be modified, including physical layer convergence protocol (PLCP) data unit (PPDU) format, spatial reuse, etc.

\subsubsection{Spatial sharing and interference mitigation}

Owing to the directional transmissions of mmWave communications, the interference is completely different from that of omni-directional communications \cite{ref19}. In both 802.11ad and 802.11ay, SPSH mechanisms were proposed to improve the throughput by allocating those STAs that do not interfere with each other to the same SP. There are more opportunities for SPSH in mmWave communications that benefit from directional transmissions and high path-loss \cite{ref19, ref20, ref21, ref22}. Therefore, how to design effective and novel methods to exploit the capabilities of SPSH is very important for interference mitigation. In Section IV, we will present some methods proposed for 802.11ay.

\subsubsection{Beamforming training and beam tracking}

BFT is a key procedure to enable directional communications. If the BFT processes take too much time to establish a directional communication link, communication delay will be too long to provide a good Quality of Experience (QoE) \cite{ref23}. However, the BFT processes in 802.11ad are fixed and cannot be configured flexibly according to different requirements which will make the BFT processes very time-consuming and inefficient in future mmWave wireless communications. Thus, efficient and low complexity BFT methods are urgently needed. Besides, beam tracking is a mechanism that allows fast link switch and recovery, and it is important for beamformed link robustness \cite{ref24}. However, the beam tracking method in 802.11ad can only try to find a better link when the operating beam becomes worse. It will suffer link blockage when a sudden interruption occurs. Unfortunately, the existing beam tracking mechanisms cannot guarantee the link robustness. Thus, efficient and intelligent beam tracking methods are highly demanded. Fortunately, there are a lot of efficient BFT and beam tracking methods proposed for 802.11ay. We will discuss some insightful methods in Section V.

\subsubsection{Beamformed link blockage}

Since mmWave frequency band suffers from serious path-loss, beamforming technologies that can concentrate TX power and RX region on a narrow beam to achieve directional communication are needed \cite{ref10}. Furthermore, the misalignment of TX beam and RX beam in mobile communication scenarios is a serious problem. Thus, simple but efficient beam tracking and BFT methods should be designed to deal with this serious issue. Once a beamformed link is blocked by obstacles such as human body, complicated and time-consuming BFT processes have to be gone through to find another available link \cite{ref25}. In order to cope with this problem, setting up relays provides a practical solution \cite{ref26}. We will elaborate a few practical solutions to relieve the blockage problems in Section V.

\subsubsection{Power management}

Power consumption is an important issue since mmWave communications shall use a large number of antenna arrays to form a directional beam \cite{ref27}. Thus, efficient power saving mechanisms should be carefully designed to search for the time periods when an STA unnecessarily stays in an active transmission mode. Although the power management mechanism described in the 802.11ad is efficient to some extent, there are still some issues to resolve. For example, 802.11ad defines a single awake window per BI, at the first CBAP allocation. If an STA is woken up in the awake window at the first CBAP, it will remain awake until the end of the BI \cite{ref28}. This is very inefficient from the power consumption perspective. For an EDMG STA involved in an MU-MIMO transmission, except for the STA with implicit Block Ack Request, all other STAs do not know when they will receive the Block Ack Request from the PCP/AP, and therefore have to power on and wait continuously until they hear the Block Ack Request from the PCP/AP \cite{ref29}. It is also inefficient for power conservation. The 802.11ay provides a better power management than the 802.11ad. More details can be found in \cite{ref4}.

\subsection{MAC related issues in the 802.11ay draft}

This subsection summarizes some important MAC related aspects of the 802.11ay \cite{ref4}, including enhanced DMG (EDMG) channel access, EDMG beamforming, SU-MIMO \& MU-MIMO operations, and SPSH and interference mitigation.

\subsubsection{EDMG channel access}

Owing to the multi-channel operational ability of 802.11ay, the A-BFT and DTI could be present on both primary channel and secondary channels \cite{ref4}. Two EDMG STAs can communicate with each other on a bonded channel or an aggregated channel to achieve higher throughput. In addition, directional allocation is also proposed for 802.11ay, which is not included in 802.11ad. Furthermore, MIMO channel access is another important feature supported in 802.11ay, which will be described in Section IV-C.

\subsubsection{EDMG beamforming}

EDMG beamforming supports BFT on a bonded channel and an aggregated channel, and antenna polarized BFT is also supported. In addition, EDMG beamforming provides methods, such as multi-channel A-BFT, extending the limited slot resources of A-BFT and spreading out the random-access attempts in dense user scenarios over the time \cite{ref4} to tackle the collision problem. BRP TXSS procedure supported in EDMG beamforming uses BRP frames to perform TXSS and determines effective antenna configuration for transmissions. Another efficient BFT method in EDMG beamforming is to append training sequences fields (TRN units) to the DMG Beacon frames. The details of the above BFT methods will be described in Section V.

\subsubsection{SU-MIMO and MU-MIMO operations}

In the 802.11ay, MIMO beamforming is another important procedure for EDMG beamforming \cite{ref4}, which can only be performed in DTI. There are two training methods, namely, SU-MIMO and MU-MIMO BFT, to enable MIMO communications between an initiator and one or more responders. Beamforming for asymmetric links and group beamforming were also proposed in the 802.11ay. We will introduce the details of beamforming for asymmetric links, MIMO beamforming and group beamforming in Section V, Section VI and Section VII, respectively.

\subsubsection{Spatial sharing (SPSH) and interference mitigation}

SPSH and interference mitigation mechanisms in 802.11ad are based on a single channel \cite{ref2}. However, this mechanism can operate on one or more channels in 802.11ay \cite{ref4}. In addition, SU-MIMO beamformed links are also supported for SPSH in 802.11ay. More details will be presented in Section IV.

In what follows, we will elaborate on the aforementioned important design issues in 802.11ay.

\section{Channel Bonding and Aggregation}
To achieve significant throughput gain, wider channels (more than 2.16 GHz) for data transmission are used in the 802.11ay. The formation of a wider channel from multiple relatively narrow channels is known as channel bonding. Actually, all the 802.11 protocols and drafts, such as the 802.11n, 802.11ac, 802.11ad and 802.11ay, etc., rarely adopt the term ``channel bonding'' formally. However, the majority of the current studies name this technology as either channel bonding or channel aggregation.

\subsection{General Description}
A bonded channel usually comprises one primary channel and one or more secondary channels. In order to guarantee the coexistence and backward compatibility with the legacy 802.11ad devices that do not support channel bonding, all control and management frames must be transmitted over a single 2.16 GHz channel, which is called the \emph{primary channel}. The rest of the channels are referred to as \emph{secondary channels}. Primary channel is the common control channel of operations for all STAs that are members of the BSS. There is only one channel serving as the primary channel and the position of the primary channel may be different. This means that the PCP/AP can dynamically choose a channel as the primary channel.

\subsection{Channel bonding}

\begin{figure}[!htbp]
  \begin{center}
    \scalebox{0.36}[0.36]{\includegraphics{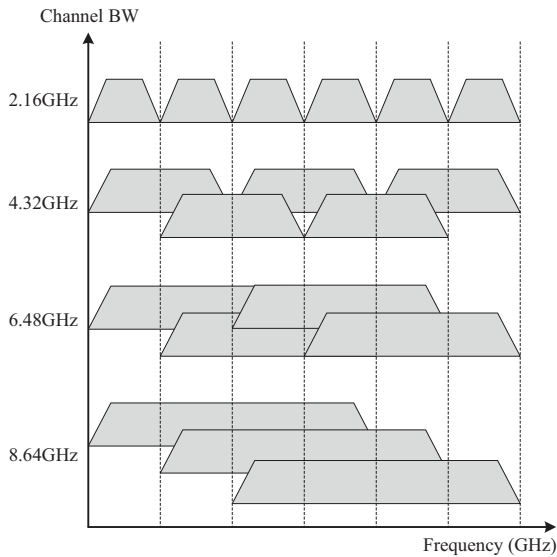}}
    \caption{Channelization used by EDMG STAs.}
  \end{center}
\end{figure}

\begin{figure*}[bp]
  \begin{center}
    \scalebox{0.68}[0.68]{\includegraphics{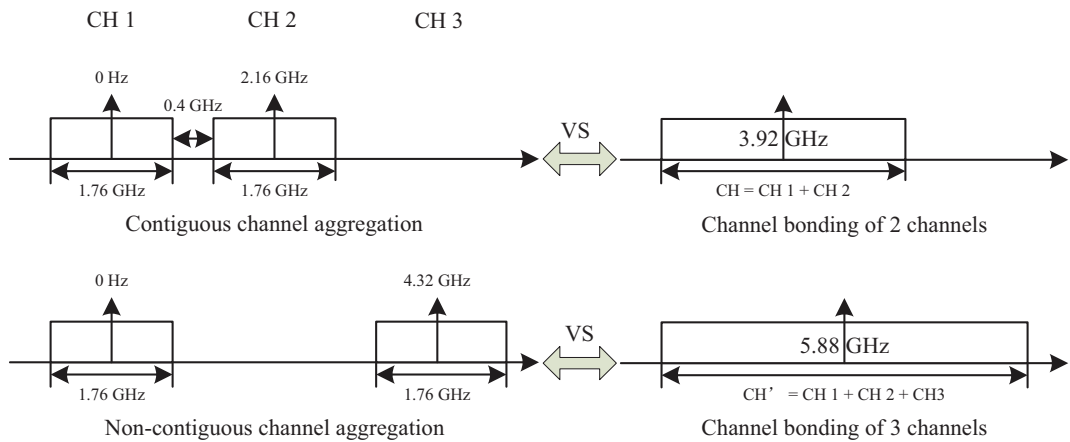}}
    \caption{Channel bonding vs channel aggregation.}
  \end{center}
\end{figure*}

\begin{table*}[bp]
\centering
\caption{Benefits and Challenges for Channel Bonding and Channel Aggregation.}
\begin{tabular}{|l|l|l|}
\Xhline{0.6pt}
~{} & \textbf{Channel bonding} & \textbf{Channel aggregation}\\        
\Xhline{0.6pt}
\textbf{Benefits} & 1) Lower PAPR in SC payload. & 1) Improve radio resource usage.\\
~{} & 2) Higher peak throughput in OFDM (about 10\% higher). & 2) Lower hardware requirements.\\
~{} & ~{} & 3) Same sampling rate as in the 802.11ad, reuse of legacy digital baseband.\\
~{} & ~{} & 4) Possibly high efficiency for mid-size packets.\\
~{} & ~{} & 5) L-STF and L-CEF can be reused for channel estimation.\\
\Xhline{0.6pt}
\textbf{Challenges} & 1) Very high-speed DAC/ADC. & 1) Adjacent Channel Interference (ACI) in contiguous channel aggregation.\\
~{} & 2) Efficient \& flexible channel usage. & 2) Allows for simultaneous sensing and detection of the two legacy \\
~{} & ~{} & ~{}~{} 802.11ad channels.\\
~{} & ~{} & 3) Co-existence with wideband channel bonding.\\
\Xhline{0.6pt}
\end{tabular}
\end{table*}

As the evolution of the 802.11ad standard, the two main features that allow 802.11ay to achieve at least 20 Gbps maximum transmission rates are as follows:

(1) wider bandwidth from multiple channels (known as channel bonding),

(2) downlink multi-user MIMO (DL MU-MIMO).

Channel bonding provides wider bandwidth and higher throughput for data transmissions, while DL MU-MIMO enables distribution of capacity to multiple STAs simultaneously. Compared with channel bonding in 802.11n/ac, the most significant difference for channel bonding in 60 GHz frequency bands lies in the directional communications and BFT. To accommodate scenarios for DL MU-MIMO and channel bonding at other mmWave frequency bands, while enabling the backward compatibility with the legacy 802.11ad devices, many design aspects, such as channel access schemes, BFT procedures, and frame formats, etc., need to be modified.

As specified in the 802.11ay \cite{ref4}, there are up to six 2.16 GHz channels and the overlapped bonded channels can be used. Fig. 3 shows the channelization used by EDMG STAs. Similar to 802.11n/ac, channels consisting of 4.32 GHz or wider always require a primary channel.

\subsection{Differences between channel bonding and channel aggregation}
It is usually confusing with channel bonding, channel aggregation and multi-channel. Few current research works clearly explain the differences among the three terms. Since both channel bonding and channel aggregation inevitably involve with multiple channels, it is obvious that multi-channel concept covers a broader range. The differences between the concepts of channel bonding and channel aggregation are subtle and more confusing. In general, the differences mainly lie in two aspects. Firstly, channel bonding usually refers to merge multiple contiguous channels into one wideband channel, and there is no channel spacing among multiple channels, which can be used as a whole hand to form a single channel. In contrast, channel aggregation is often used as the combination of two or more contiguous or non-contiguous channels, and there exists channel spacing among these channels. Fig. 4 shows the difference between channel bonding and channel aggregation. The other different aspect between channel bonding and channel aggregation is the physical frame format \cite{ref4}.

Generally speaking, both channel bonding and channel aggregation belong to multi-channel operations. The benefits and challenges were analyzed in \cite{ref30, ref31, ref32}, as shown in Table I.

\section{Channel Access and Allocation}

\subsection{General Overview}
In 802.11ad, it allows scheduled channel access in which an STA knows the start time and the duration of its scheduling period in advance to obtain higher QoS. However, the 802.11ad standard does not support multi-channel operations as we described before. It lacks the use of plenty of frequency resources and large available spectrum stocked in 60 GHz mmWave band. Therefore, the channel access and SPSH in the 802.11ay are enhancements of the existing 802.11ad standard which support the multi-channel operations and MIMO operations. Multi-channel operations can significantly enhance channel utilization efficiency. Channel bonding and channel aggregation have been considered to improve channel utilization and maximize multiplexing flexibility because multiple channels can be used simultaneously as described in Section III. MIMO channel access supports simultaneous transmission and reception beams through multiple DMG antennas, which therefore can significantly improve the spatial reuse and the overall throughput.

\subsection{Channel access and allocation over multiple channels}
In 802.11ad, all of the channel access periods within a BI, such as BTI, A-BFT, ATI and DTI are performed over a single channel. DTI is coordinated as CBAP or SP using a schedule by the PCP/AP. In 802.11ay, the EDMG PCP/AP can allocate multiple EDMG STAs on different channels to communicate with the EDMG PCP/AP simultaneously, and two EDMG STAs can communicate with each other on a bonded channel or an aggregated channel to achieve higher throughput and improve channel utilization. The current 802.11ay draft specifies the channel access rules over multiple channels as follows \cite{ref4}:

(1) The BTI and ATI shall only be present on the primary channel of the BSS.

(2) The A-BFT shall be present not only on the primary channel of the BSS, but also be present on adjacent secondary channel of the BSS.

(3) Transmissions shall not occupy a bandwidth that exceeds the equivalent bandwidth of four 2.16 GHz channels.

(4) The PCP/AP can schedule an allocation of SP(s) and scheduled CBAP(s) over more than one channel and over a bonded channel. That is to say, CBAP that follows carrier sense multiple access with collision avoidance (CSMA/CA) and SP that is allocated by time division multiple access (TDMA) mode shall expand to multiple channels \cite{ref33}. Note that the allocation does not have to include the primary channel. If the allocation does not include the primary channel, the allocation shall not span more than one 2.16 GHz channel.

In order to improve the channel utilization and maximize the multiplexing flexibility for 802.11ay, a general procedure of frequency multiple access was proposed in \cite{ref34}. As shown in Fig. 5, the PCP/AP that has the capability of multi-channel operations transmits data through wider bandwidth and allocates each STA to different channels for frequency multiplexing, respectively. The intended STAs receive data through allocated channels by decoding the 802.11ay PPDU. In this way, multiple users can access the medium in multiple channels within a transmission opportunity (TXOP) and thus enhancement on channel utilization efficiency through allocating multiple users to multiple channels can be achieved.

\begin{figure}[!htbp]
  \begin{center}
    \scalebox{0.43}[0.43]{\includegraphics{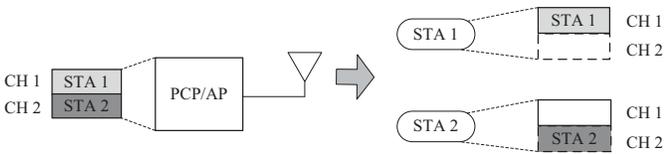}}
    \caption{Frequency multiple access procedure.}
  \end{center}
\end{figure}

Since the 802.11ay supports allocations of SP(s) and scheduled CBAP(s) over more than one channel and/or over a bonded channel \cite{ref35}, in order to improve channel utilization, an efficient multi-channel operation was proposed in \cite{ref36} to allow only secondary channel allocation. As depicted in Fig. 6, when the allocation \#1 is allocated to legacy device and the allocation \#2 is allocated to the 802.11ay devices, the 802.11ay devices can use the secondary channel (CH 2) regardless of the state of the primary channel (CH 1). Note that the allocation \#1 and \#2 can be a CBAP or SP. The whole procedure requires that the EDMG STAs be allocated to only a secondary channel by the PCP/AP which transmits DMG Beacon and announce frames through the primary channel. Notice that all the EDMG STAs are working on the primary channel except for the allocation periods. Besides, the collision problems are also considered. If the type of allocation on a secondary channel is CBAP, EDMG STAs shall operate backoff procedure before getting the channel access. Otherwise, if the type of allocation on a secondary channel is SP, EDMG STAs shall execute DMG protected SP access operation similar to the 802.11ad. In both cases, the collision problems can be alleviated effectively.

\begin{figure}[!htbp]
  \begin{center}
    \scalebox{0.43}[0.43]{\includegraphics{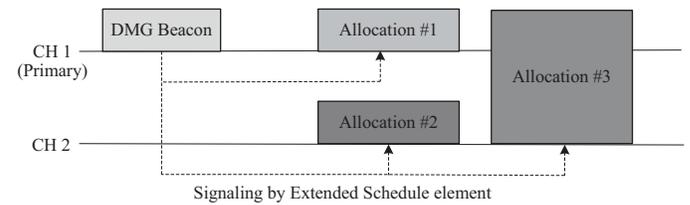}}
    \caption{A secondary channel allocation scheme.}
  \end{center}
\end{figure}

\begin{figure*}[!htbp]
  \begin{center}
    \scalebox{0.85}[0.85]{\includegraphics{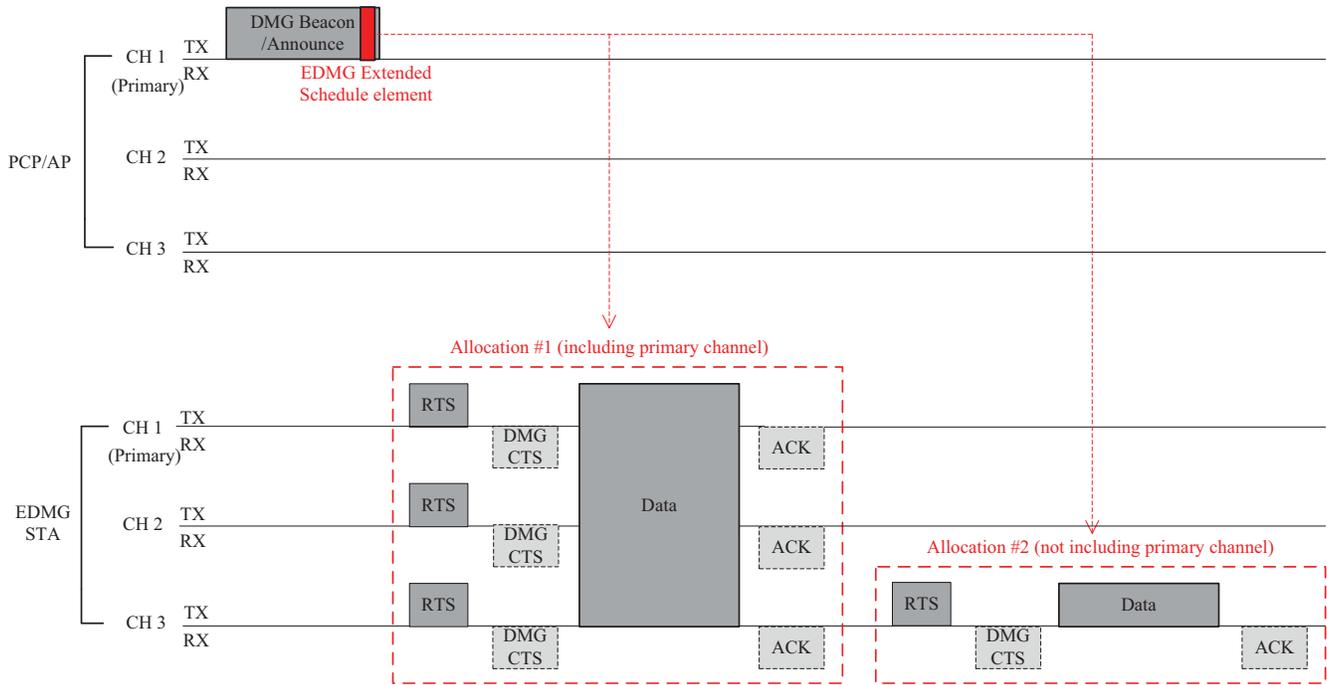}}
    \caption{Full carrier sensing applied to an allocation that does not include primary channel.}
  \end{center}
\end{figure*}

If the allocation does not include the primary channel \cite{ref37}, there will be an issue about the carrier sensing mechanism for multi-channel allocation. A carrier sensing scheme for multi-channel allocation to tackle the above problem was proposed in \cite{ref38}. Since the carrier sensing mechanism is not clear for the allocation that does not include primary channel and if only physical carrier sensing is applied to this allocation, the EDMG STAs allocated to a non-primary channel cannot set the network allocation vector (NAV), and thus it may increase the collision probability in overlapping BSS (OBSS) environments. It is suggested to perform full carrier sensing (physical and virtual) on the channel that does not include the primary channel during the allocation process. The virtual carrier sensing can efficiently avoid the hidden node problem as shown in Fig. 7. The physical carrier sensing (i.e., Clear Channel Assessment, CCA) includes energy detection and carrier sensing provided by the PHY layer. The virtual carrier sensing consists of a Request-to-Send (RTS)-DMG Clear-to-Send (CTS) handshake procedure. These two carrier sensing mechanisms are similar to the legacy 802.11 standard, but are applied to the channel that does not include the primary channel.

In the existing 802.11 protocol, the TXOP must be obtained by each STA on its primary channel and that cannot fully and efficiently utilize the spectral resources in a frequency band if congestion occurs in a channel. To resolve this problem, Yano \emph{et al.} \cite{ref39} first defined an alternative primary channel (APCH) where each STA is able to obtain TXOP to transmit if it is idle for some time, even when the primary channel is busy. Then, a channel access procedure was proposed to implement load balancing of channel access among multiple channels.

\subsection{MIMO channel access}
MIMO channel access is another important feature supported by 802.11ay. An EDMG STA with multiple TX antennas can transmit multiple streams to a peer EDMG STA with multiple RX antennas. On the one hand, it can improve the spatial reuse and provide a highly efficient transmission. On the other hand, it can boost the robustness by avoiding the outage when one of the streams is blocked as shown in Fig. 8 \cite{ref40}.

\begin{figure}[!htbp]
  \begin{center}
    \scalebox{0.34}[0.34]{\includegraphics{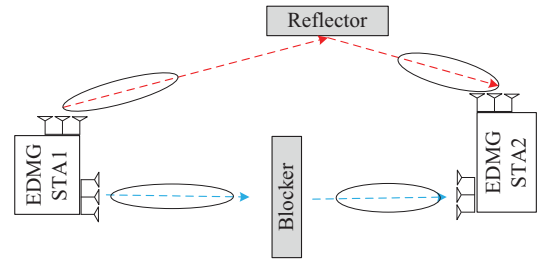}}
    \caption{SU-MIMO transmission scenario.}
  \end{center}
\end{figure}

MIMO channel access for 802.11ay was discussed in \cite{ref41}. It assumes that MIMO initiator and MIMO responder use RTS and DMG CTS frames to exchange channel availability and to set NAV on other STAs. It is suggested that in all cases, a device shall be able to receive single input single output (SISO) frames when MIMO channel access is pending. It provides two options for MIMO channel access and both options require the support of SISO reception.

(1) Physical carrier sensing and virtual carrier sensing shall be maintained such that at least all the beams used for MIMO communications are covered, and it has two ways. The one is to create a new MIMO backoff timer. If all the MIMO beams have channel idle, the MIMO backoff timer decreases. If not, all the MIMO beams have channel idle, the MIMO backoff timer freezes. MIMO channel access is allowed when the MIMO backoff timer reaches zero. The other is that one or more backoff timers may be used for MIMO channel access, and each backoff timer decreases when the associated physical carrier sensing and virtual carrier sensing shows that the channel is clear, and remains the same when the channel is busy. MIMO channel access is allowed when at least one backoff timer reaches zero and all the rest of the backoff timers are not suspended.

(2) At least CCA of energy detection shall be maintained such that at least all the beams that are used for MIMO communications are covered. When the backoff timer reaches 0, and all the beams used for MIMO communications have CCA cleared for Point Coordination Function Interframe Space (PIFS), MIMO channel access is allowed.

In view of the complexity and power consumption, option 2 is recommended. The procedure of SU-MIMO channel access was described in \cite{ref3}.

Considering the fact that partial successes among simultaneous transmissions can cause a severe unfairness problem because channel access opportunity may occur among STAs with different PER, Cha \emph{et al.} in \cite{ref42} proposed a novel MAC protocol which conditionally increases the contention window size on the basis of the number of simultaneous transmissions to guarantee fair channel access for STAs in uplink (UL) WLANs with MU-MIMO.

\subsection{Spatial sharing (SPSH) and interference mitigation}
The SPSH mechanism allows SPs belonging to different DMG STAs in the same spatial vicinity to be scheduled concurrently over the same channel in the 802.11ad standard \cite{ref2}. The DMG STAs involved in a candidate SP need to perform measurements before achieving SPSH. The DMG STAs involved in existing SP also need to perform measurements if possible. Before conducting measurements, DMG STAs involved in a candidate SP and the existing SP need to be beamforming trained with each other. A DMG STA shall carry out the measurement by employing the same RX antenna configuration as being used when receiving the single spatial stream from the target DMG STA.

The 802.11ay supports SU-MIMO operations, which can significantly improve the spatial reuse capability. Thus, an EDMG STA with multiple TX antennas can transmit multiple streams to a peer EDMG STA with multiple RX antennas. A DMG antenna is typically used with a single antenna configuration at a time for the transmission or the reception of a stream. Multiple TX/RX antenna configuration combinations for SU-MIMO operations can be obtained via MIMO BFT as described in Section VI-B. Under this circumstance, it is necessary to consider the existing or a candidate SP with SU-MIMO transmissions.

The EDMG SPSH mechanism enables an EDMG STA to perform concurrent measurements employing multiple RX antenna configurations as being used for receiving multiple streams from the target EDMG STA based on the same measurement configurations \cite{ref43}. As shown in Fig. 9, the EDMG PCP/AP transmits an enhanced Directional Channel Quality Request to an EDMG STA to request it to conduct multiple measurements concurrently by employing multiple RX antenna configurations as being used for receiving multiple streams from the target EDMG STA. Each of multiple RX antenna configurations corresponds to a specific RX antenna. In addition, two measurement reporting methods after performing concurrent measurements were proposed. The first method was to report the results of concurrent measurements during each measurement time block individually. The second one was to report the average of the results of concurrent measurements during each measurement time block. Compared with the first method, the second method has a much shorter measurement report. However, the first method can provide more detailed measurements. The detailed modifications on the frame formats can be found in \cite{ref43, ref44, ref45}.

\begin{figure}[!htbp]
  \begin{center}
    \scalebox{0.42}[0.42]{\includegraphics{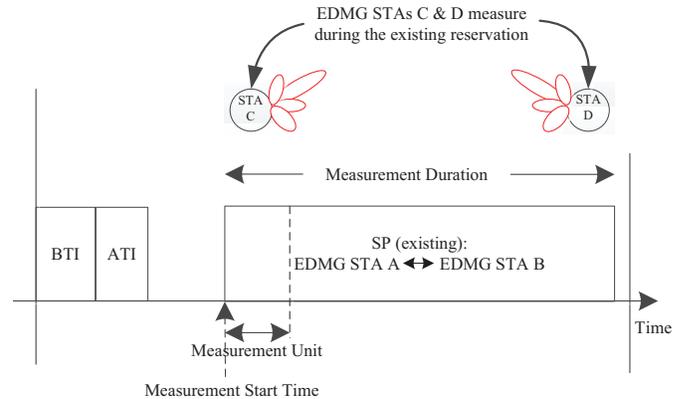}}
    \caption{Concurrent measurements over multiple RX antenna configurations.}
  \end{center}
\end{figure}

A multi-channel SPSH scheme to reinforce the existing SPSH protocol was proposed in \cite{ref46}. The main idea is to modify the Directional Channel Quality Request/Report frames for multi-channel SPSH. It adds a Measurement channel bitmap for the channels requested by AP in the reserved bits of the Directional Channel Quality Request frame and for the channels measured by STAs in the reserved bits of the Directional Channel Quality Report frame to keep backward compatibility with the legacy systems. In addition, it also adds two methods to report channel measurements. The first method is to collect results of concurrent measurements individually while the second method is to get the average of the results of concurrent measurements as we discussed earlier. The detailed modifications on the frame format can be found in \cite{ref45, ref46, ref47}.

In addition, although many research works have drawn attention to the SPSH in mmWave communications, few of them deal with inter-PBSS SPSH within a cluster. Thus, a method of SPSH with interference mitigation among multiple co-channel PBSSs was proposed in \cite{ref22}. Within a cluster, each PCP/AP obtains information of the co-existing links in other PBSSs that may cause interference through training and exchanging a newly defined SPSH report. Thus, the Inter-PBSS SPSH mutual avoidance algorithm is proposed to make full use of SPSH between different PBSSs, thus improving network throughput.

\section{Efficient Beamforming Training and Beam Tracking}

\subsection{General Overview}

The basic procedures of beamforming in the 802.11ay are almost the same as those in the 802.11ad standard (as summarized in Section II-A), but 802.11ay provides more flexible and efficient beamforming methods than the 802.11ad with the enhancement and optimization.

\begin{figure*}[!htbp]
  \begin{center}
    \scalebox{0.82}[0.82]{\includegraphics{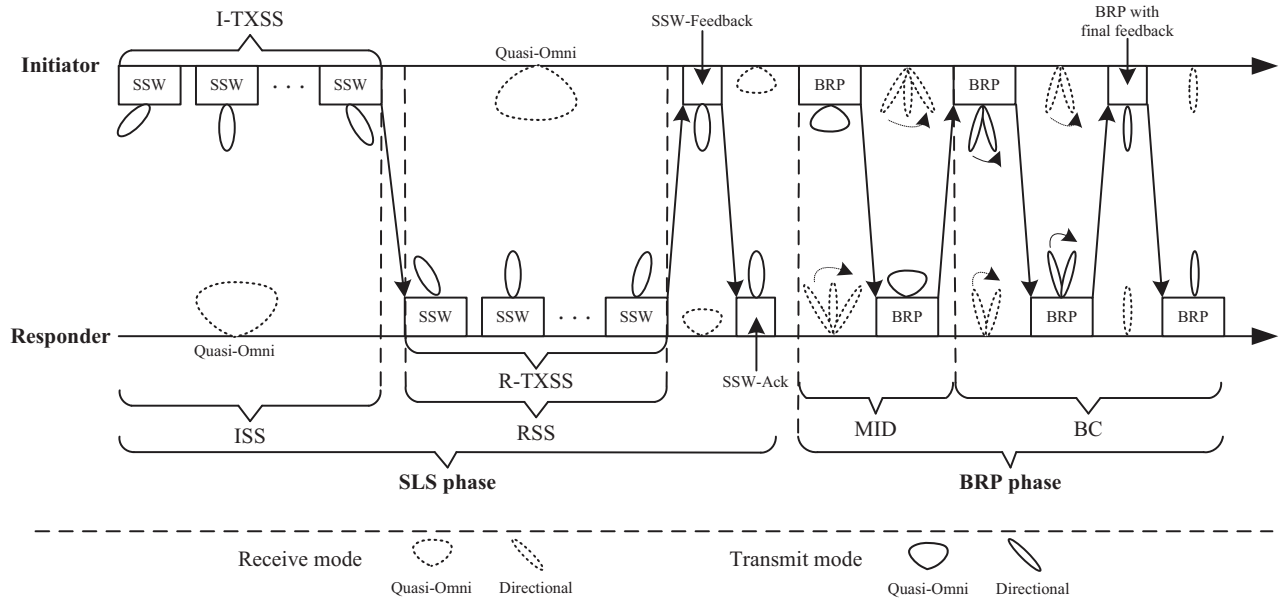}}
    \caption{An overview of beamforming training.}
  \end{center}
\end{figure*}

\begin{figure*}[!htbp]
  \begin{center}
    \scalebox{0.63}[0.63]{\includegraphics{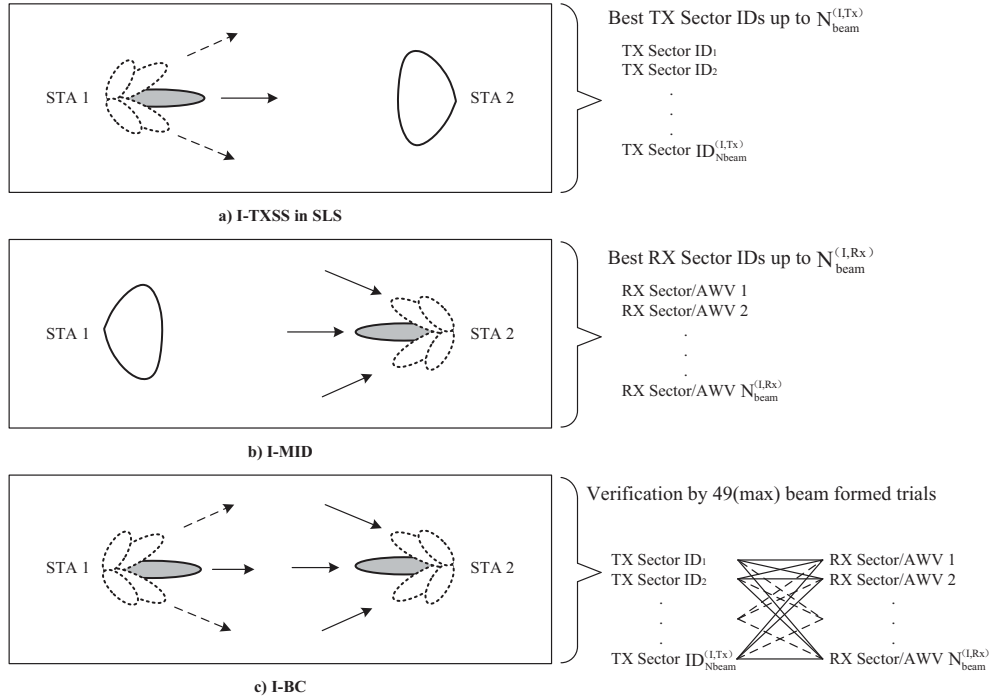}}
    \caption{An example of SLS, MID and BC for the initiator link.}
  \end{center}
\end{figure*}

Fig. 10 gives an overview of a basic BFT procedure. The SLS phase typically contains four subphases, which are initiator sector sweep (ISS), responder sector sweep (RSS), sector sweep feedback (SSW-FBCK), and sector sweep ack (SSW-Ack). It is noteworthy that the SLS phase only trains TX sectors of both initiator and responder and is mandatory, while the BRP is optional. If either initiator or responder wants to train RX sectors and obtain refined antenna weight vectors (AWVs) of both transmitter and receiver, the BRP phase will follow the SLS. More details are given as follows.

\subsubsection{Sector-level sweep (SLS) phase}

\begin{figure*}[!htbp]
  \begin{center}
    \scalebox{0.67}[0.67]{\includegraphics{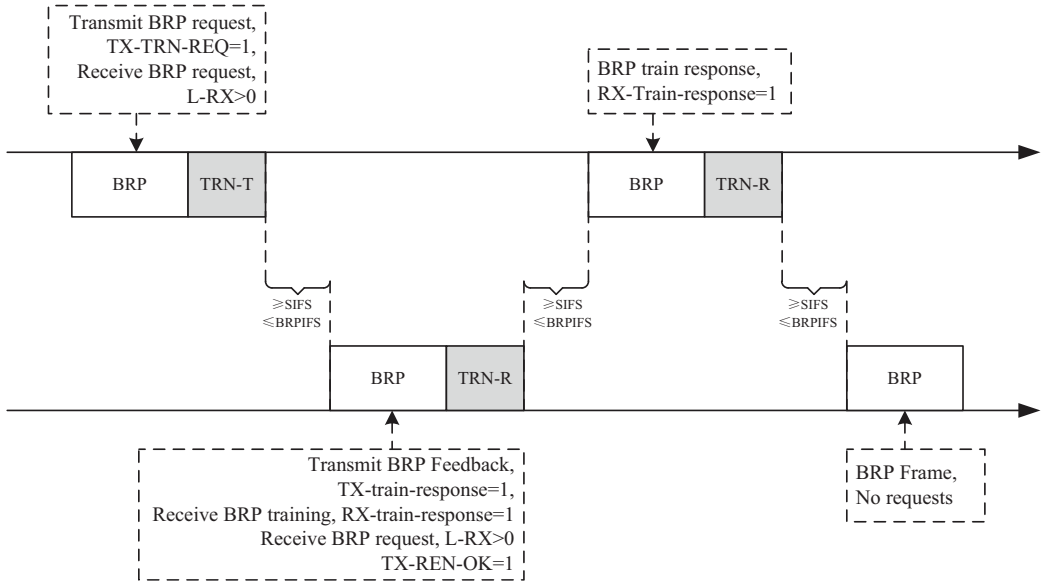}}
    \caption{An example of beam refinement transaction.}
  \end{center}
\end{figure*}

As seen from Fig. 10, the initiator starts with the SLS and transmits DMG Beacon or SSW frames (or Short SSW frames for 802.11ay) to train its TX sectors, and then the responder transmits SSW frames (or Short SSW frames for the 802.11ay) to train its TX sectors. The SSW frames (or Short SSW frames for 802.11ay) transmitted by responder contain the best TX sector ID of the initiator obtained from the ISS subphase. Then, the initiator sends backs the best TX sector ID of the responder obtained from the RSS with its best TX sector. After the responder receives SSW-FBCK frame, it will use the best TX sector to transmit an SSW-Ack frame to the initiator. When the SLS of beamforming is completed, the communications between the two participating STAs with the DMG control mode rate or higher modulation and coding scheme (MCS) are enabled.

\subsubsection{Beam Refinement Protocol (BRP) phase}
Since the TX sectors of both initiator and responder have been trained during the SLS phase, the RX sectors of both initiator and responder have not been trained yet. The BRP phase aims to train the RX sectors and obtain refined AWVs. It typically contains BRP setup subphase, MID subphase, BC subphase, and beam refinement transactions subphase. The intent and capabilities of these subphases are exchanged in the BRP setup subphase through the BRP frames. As shown in Fig. 11, in the MID subphase, the initiator transmits BRP frames in a quasi-omni direction, and the responder trains its RX sectors with directional reception mode. After the SLS and MID phases, the TX and RX sectors of both the initiator and the responder have been trained. To find the best beamformed link, TX sectors and RX sectors should be paired. In order to reduce the sector pairing time, a limited number of TX sectors and RX sectors should be selected. The optimal beam pair should be considered as the communication link, while the suboptimal beam pairs can be used as the backup in case of the optimal communication link being interrupted.

The beam refinement transaction subphase is used to explore a broader set of TX and RX AWVs with the help of exchanging request and response frames. As observed from Fig. 12, both the initiator and the responder can append TRN units at the end of BRP frames to train its TX sectors or RX sectors. More specifically, STA can use transmit training (TRN-T) units to train its TX sectors and TRN-R units to train its RX sectors.

\subsection{Beamforming in BTI}
In BTI, the initiator transmits DMG Beacon frames to perform the I-TXSS. Notice that the SSW frames and the Short SSW frames cannot be transmitted in BTI as they have limited functions. The 802.11ay provides an efficient receive BFT method during BTI by using TRN-R units appended to the end of DMG Beacon frames. As shown in Fig. 13, while an EDMG STA receives this kind of DMG Beacon frame, it will stay in the quasi-omni receiving mode to receive the DMG Beacon frame, and then operate in the directional receiving mode to train its RX sectors among the TRN-R units. By adopting TRN-R units during BTI, both initiator's TX sectors and responder's RX sectors are trained in the BTI phase. Therefore, the efficiency of BFT is improved and the time consumed for BFT is reduced.

\begin{figure}[!htbp]
  \begin{center}
    \scalebox{0.42}[0.42]{\includegraphics{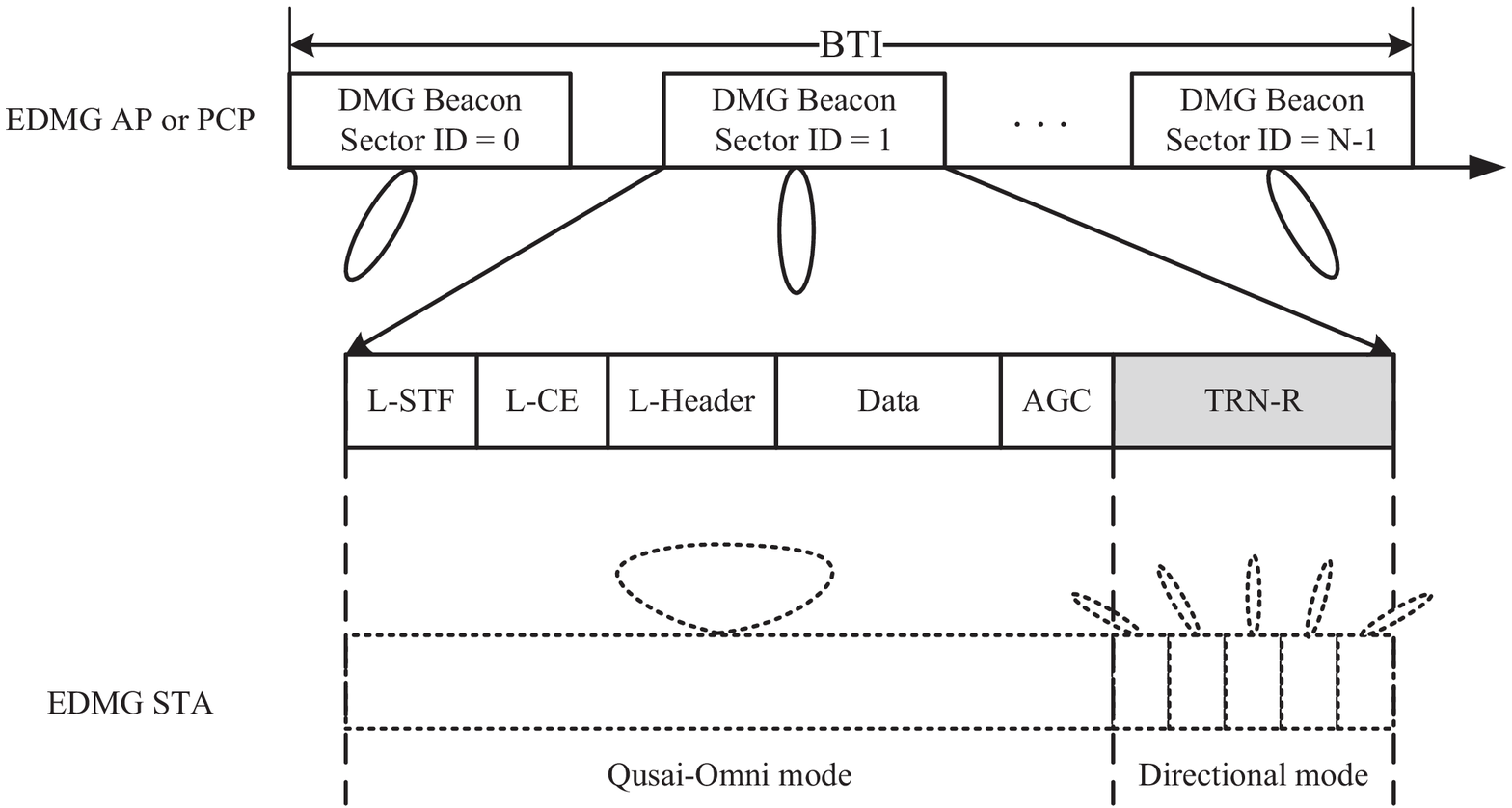}}
    \caption{Receive beamforming training during BTI by using TRN-R units.}
  \end{center}
\end{figure}

\begin{figure}[!htbp]
  \begin{center}
    \scalebox{0.8}[0.8]{\includegraphics{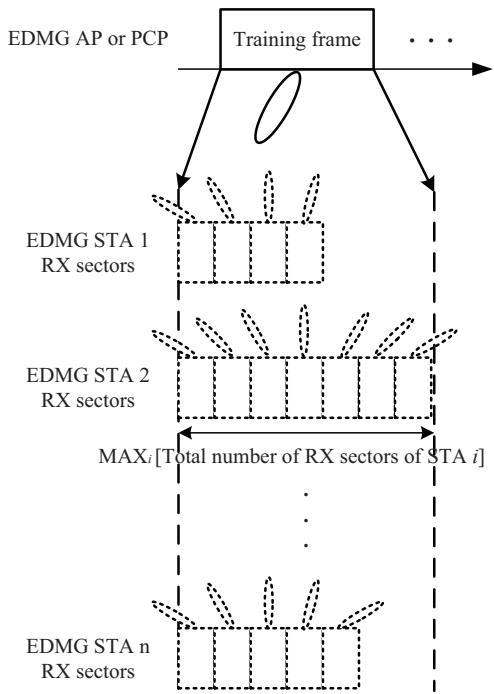}}
    \caption{Scalable beamforming training in BTI.}
  \end{center}
\end{figure}

\begin{figure}[!htbp]
  \begin{center}
    \scalebox{0.43}[0.43]{\includegraphics{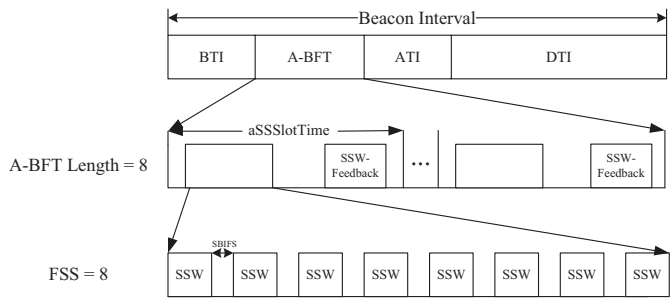}}
    \caption{A-BFT structure.}
  \end{center}
\end{figure}

\begin{figure*}[bp]
  \begin{center}
    \scalebox{0.86}[0.86]{\includegraphics{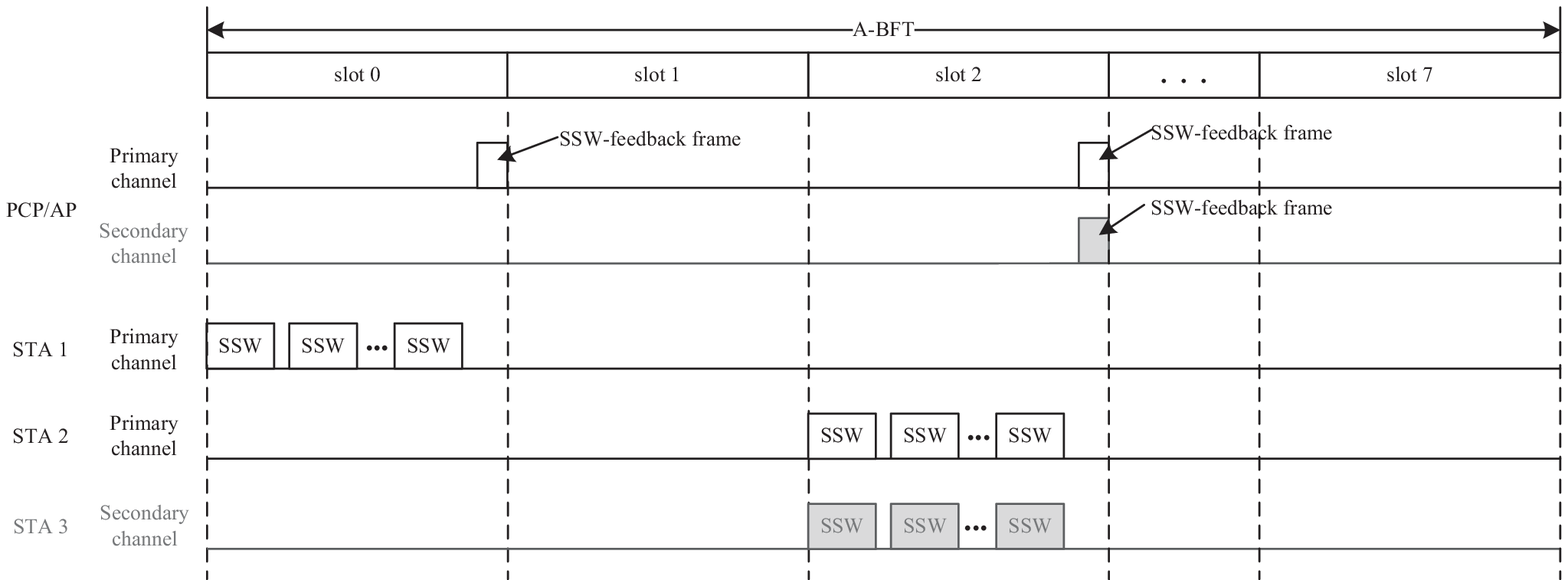}}
    \caption{RSS in A-BFT over multiple channels.}
  \end{center}
\end{figure*}

A scalable BFT method was proposed in \cite{ref48}, where it adopted antenna pattern reciprocity to reduce BFT overhead. As shown in Fig. 14, it can achieve one-to-multiple STAs BFT simultaneously. The responders do not need to perform the responder R-TXSS since they can derive results from their receive BFT by utilizing the antenna pattern reciprocity. When an initiator (the EDMG PCP/AP) transmits BFT frames (e.g., DMG Beacon frame, SSW frame, Short SSW frame, BRP frame, etc.), multiple responders (EDMG STAs) can receive simultaneously. Thus, the BFT time is determined by the number of initiator's TX sectors * $\max \limits_i$[number of responder \emph{i}'s RX sectors], which is independent of the number of responders.

\subsection{Beamforming in A-BFT}
Following the end of BTI, RSS and SSW feedback will be performed in the A-BFT subphase between the PCP/AP and STAs. As seen from Fig. 15, the A-BFT consists of several slots (up to 8 slots in the 802.11ad), and the number of slots is indicated by the A-BFT Length field of DMG Beacon frame. After receiving DMG Beacon frame from BTI, STAs randomly select a slot to perform RSS by using SSW frames or Short SSW frames. If a slot selected by two or more STAs to perform RSS, collision will occur and this slot may fail for RSS. Thus, the PCP/AP will not transmit SSW-FBCK frame in this slot. The STAs failed RSS in this A-BFT will try to perform RSS again in the next BI's A-BFT subphase.

In \cite{ref49}, the performance of A-BFT in the 802.11ad was evaluated. Since there is only one channel supported in the 802.11ad and STAs will randomly select a A-BFT slot to perform RSS, A-BFT suffers from serious collisions as the number of STAs involved in A-BFT increases. However, 802.11ay supports multi-channel operations, and therefore, Xin \emph{et al.} proposed a multi-channel A-BFT scheme in \cite{ref50} to reduce the serious collision during A-BFT in dense user scenarios. As shown in Fig. 16, A-BFT allocated over the secondary channels has the same configuration as the A-BFT allocated on the primary channel. If two or more STAs (e.g., STA 2 and STA 3) select the same slot, but they are allocated on different channels (STA 2 on primary channel, STA 3 on secondary channel), collision can be avoided.

The 802.11ad defines two parameters, namely, dot11RSSRetryLimit and dot11RSSBackoff. If the number of times an STA fails to perform A-BFT exceeds dot11RSSRetryLimit, it will retry A-BFT after a period of time equal to dot11RSSBackoff. It is neither flexible and nor efficient if there are a large number of STAs joining the BSS simultaneously, since the majority of them will exceed dot11RSSRetryLimit at the same time and they will wait for the same time (e.g., dot11RSSBackoff) to retry A-BFT. Collisions will be certainly serious in this case, since both dot11RSSRetryLimit and dot11RSSBackoff are fixed values. To overcome this shortcoming, three parameters, namely, FailedRSSAttempts, RSSRetryLimit and RSSBackoff, were defined in \cite{ref51}. Each STA maintains a counter, FailedRSSAttempts, indicating the failed times in A-BFT. For a DMG STA in the 802.11ad, RSSRetryLimit = dot11RSSRetryLimit and RSSBackoff = dot11RSSBackoff. However, an EDMG STA in the 802.11ay randomly picks the numbers in [0, RSSRetryLimit - 1) and [0, RSSBackoff - 1). Therefore, the collision can be relieved when there are a large number of STAs joining the BSS simultaneously.

Since there are at most 8 slots in A-BFT, another straightforward way to reduce collisions in dense user scenarios is to extend the number of slots. \cite{ref49} provided a separated A-BFT (SA-BFT) mechanism to extend the legacy slots to more slots and maintain backward compatibility with the 802.11ad standard. As shown in Fig. 17, an E-A-BFT Length field was defined in the DMG Beacon frame to indicate the number of additional slots extended for EDMG STAs, and the start point of the additional slots is set to A-BFT Length. Thus, the total number of slots in A-BFT is `A-BFT Length + E-A-BFT Length'. Since the start time of ATI subphase is indicated by the `Start Time field' in the Next DMG ATI element of DMG Beacon frame, in order to keep compatibility with 802.11ad, we can adjust the `Start Time field' to a time equal to `A-BFT Length + E-A-BFT Length' rather than `A-BFT Length'. In this way, DMG STAs and EDMG STAs randomly select slots from [slot 0, slot `A-BFT Length') and [slot `A-BFT Length', slot `A-BFT Length + E-A-BFT Length') to perform RSS, respectively. In addition, EDMG STAs can use Short SSW frames to perform RSS in [slot `A-BFT Length', slot `A-BFT Length + E-A-BFT Length') as well. Based on the SA-BFT mechanism, \cite{ref49} further proposed a secondary backoff A-BFT (SBA-BFT) mechanism to reduce the collision in ultra-dense user scenarios. The detailed design is referred to \cite{ref49}.

\begin{figure}[!htbp]
  \begin{center}
    \scalebox{0.43}[0.43]{\includegraphics{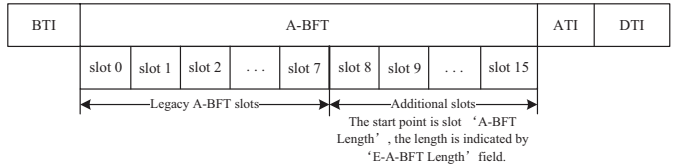}}
    \caption{SA-BFT mechanism.}
  \end{center}
\end{figure}

A similar method compared with \cite{ref49} was proposed in \cite{ref52} to extend the slots of A-BFT. As shown in Fig. 18, the additional slots are inserted before legacy slots. It can also maintain compatibility with the 802.11ad standard by adjusting the Duration field in the DMG Beacon to accommodate additional slots. Although DMG STAs cannot recognize new slots, they will wait Duration to start A-BFT.

\begin{figure}[!htbp]
  \begin{center}
    \scalebox{0.44}[0.44]{\includegraphics{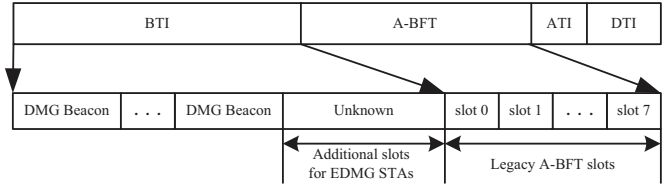}}
    \caption{Extend A-BFT slots before the legacy A-BFT.}
  \end{center}
\end{figure}

The A-BFT is the bottleneck of channel access and BFT, since high collision probability occurs in dense user scenarios. Once an STA fails to perform RSS in A-BFT, it will not be assigned to DTI and should wait for another A-BFT in the next BI, which will affect the QoE. Fortunately, collision problem of A-BFT can be significantly alleviated by SA-BFT and SBA-BFT schemes proposed in \cite{ref49}, and other schemes such as \cite{ref51} and \cite{ref52}. Since the A-BFT is slotted, when one STA performs RSS in a certain slot, other STAs may stay idle. To make full use of the A-BFT, Akhtar and Ergenthe \cite{ref53} proposed an Intelligent Listening during A-BFT (ILA) mechanism to reduce the overhead of BFT between STAs. While one STA performs RSS with AP during A-BFT, other STAs listen in the quasi-omni mode instead of staying idle and they will get to know the approximate beam direction of the training STA through the `Sector ID field' contained in the training frames. After gathering enough information, the BFT overhead between STAs can be significantly reduced since one STA can infer the approximate direction of its peer STAs and the STA does not need to search beams within all 360 degree range.

\subsection{Beamforming in DTI}
The DTI mainly contains BRP phase and data transmission. The basic processes of the BRP phase in 802.11ay have been introduced in Section V-A. As described previously, a DTI consists of several SPs and CBAPs while SPs can be allocated to perform BFT. There are many enhanced BFT schemes proposed by several companies during the process of designing the 802.11ay.

Technologies such as antenna polarization have been employed in 802.11ay to improve the performance of MIMO systems in line-of-sight (LOS) scenarios. Thus, ``multi-BF'', which uses simultaneous multi-sector steering in the polarized directions, was proposed in \cite{ref54} to reduce the time for BFT as shown in Fig. 19.

\begin{figure*}[!htbp]
  \begin{center}
    \scalebox{0.86}[0.86]{\includegraphics{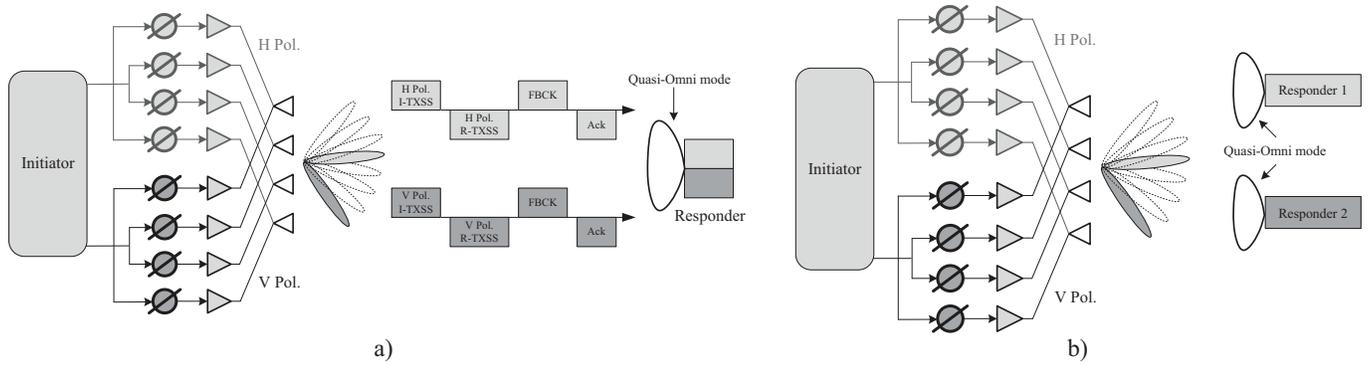}}
    \caption{ a) Simultaneous beam-steering for an STA with dual polarization. b) Simultaneous beam-steering for an STA with single polarization.}
  \end{center}
\end{figure*}

\begin{figure*}[!htbp]
  \begin{center}
    \scalebox{0.86}[0.86]{\includegraphics{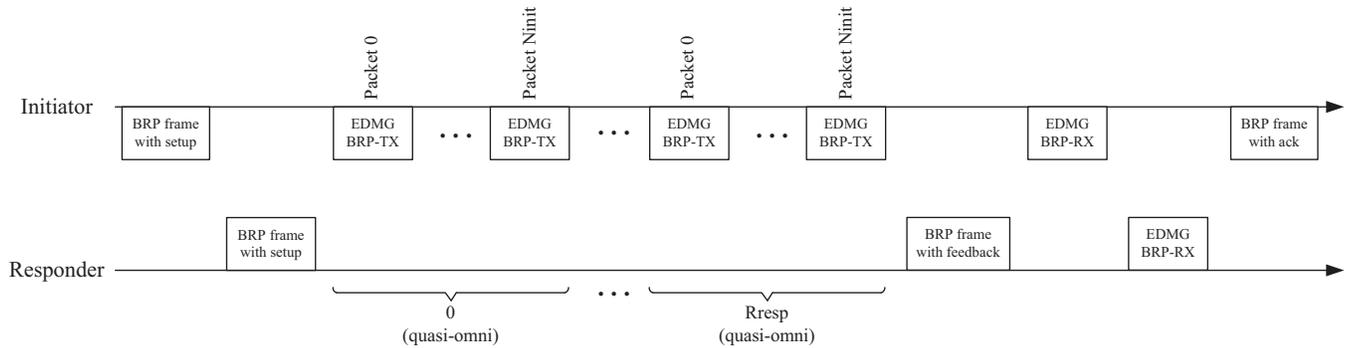}}
    \caption{An example of BRP TXSS.}
  \end{center}
\end{figure*}

In Fig. 19, the initiator/responder transmits/receives the BFT frames (e.g., the SSW frame or the BRP frame) with horizontal and vertical polarization domain simultaneously. It can reduce the time for BFT when half of the total BFT frames are transmitted by H-pole and the remaining frames are sent by V-pole separately.

A sector sweep BFT protocol, which uses BRP frames, was proposed in \cite{ref55} and presented in Fig. 20. It is efficient in terms of time, which is critical for STAs equipped with multiple antennas/subarrays. The initiator and responder exchange BRP frame to start BRP TXSS BFT. Then, the initiator transmits EDMG BRP-TX packets to perform TXSS using each of its DMG antennas, and the process is repeated for each DMG antenna of the receiver. The BRP frame with feedback contains feedback of the corresponding procedure based on measurements carried out by the responder during the reception of EDMG BRP-TX packets. This protocol can be applied for TXSS and RXSS BFT of both initiator and responder.

A multi-resolution BFT framework was introduced in \cite{ref56} and 802.15.3c \cite{ref8} in which the initiator or the responder can request BFT at a particular resolution level. Therefore, a dynamical trade-off between BFT efficiency and beam accuracy can be carried out. Another benefit of the multi-resolution BFT is that, without the need to sweep through all the beams, the beamformed link between initiator and responder can be obtained.

\subsection{Beamforming for channel bonding and channel aggregation}
As described in Section III, channel bonding and channel aggregation are the key features in 802.11ay. A BFT scheme for channel bonding and channel aggregation by using BRP frames was proposed in \cite{ref57}. Owing to the strong LOS channel characteristics in mmWave frequency band, it is unnecessary to perform a complete SLS or BRP phase over a bonded channel or an aggregated channel. Instead, we can append TRN unit that is efficient and flexible for beam refinement to BRP frames. As shown in Fig. 21, three kinds of BRP PPDUs were proposed in \cite{ref57} to cope with the following three cases. The first case is that an initiator has trained beamforming on a bonded channel with a responder. The second one is that an initiator has not trained beamforming on a bonded channel with a responder. The last one relies on an initiator that has not trained beamforming on an aggregated channel with a responder. For the first case (as depicted in Fig. 21 a)), assuming that the best sector is sector \emph{m} in both channel 1 and channel 2 from previous BFT results, the duplicate part of PPDU can be decoded through channel 1 (primary channel), the bonded part of PPDU can be decoded through the bonded channel, and the training part of PPDU conducts beam refinement for a bonded channel by using TRN units. The PPDU format as shown in Fig. 21 b) is suitable for the BFT for channel bonding when an initiator has not trained beamforming on a bonded channel with a responder. The duplicate part of PPDU can also be decoded through channel 1 (primary channel) and the best sector for the bonded channel cannot be found until conducting the training part of PPDU. The third case is similar to the second one, and the best sector for the aggregated channel cannot be found until conducting the training part of PPDU. By adopting the proposed BRP PPDUs to BFT for channel bonding and channel aggregation, the complicated and time-consuming BFT processes can be avoided.

\begin{figure*}[!htbp]
  \begin{center}
    \scalebox{0.6}[0.6]{\includegraphics{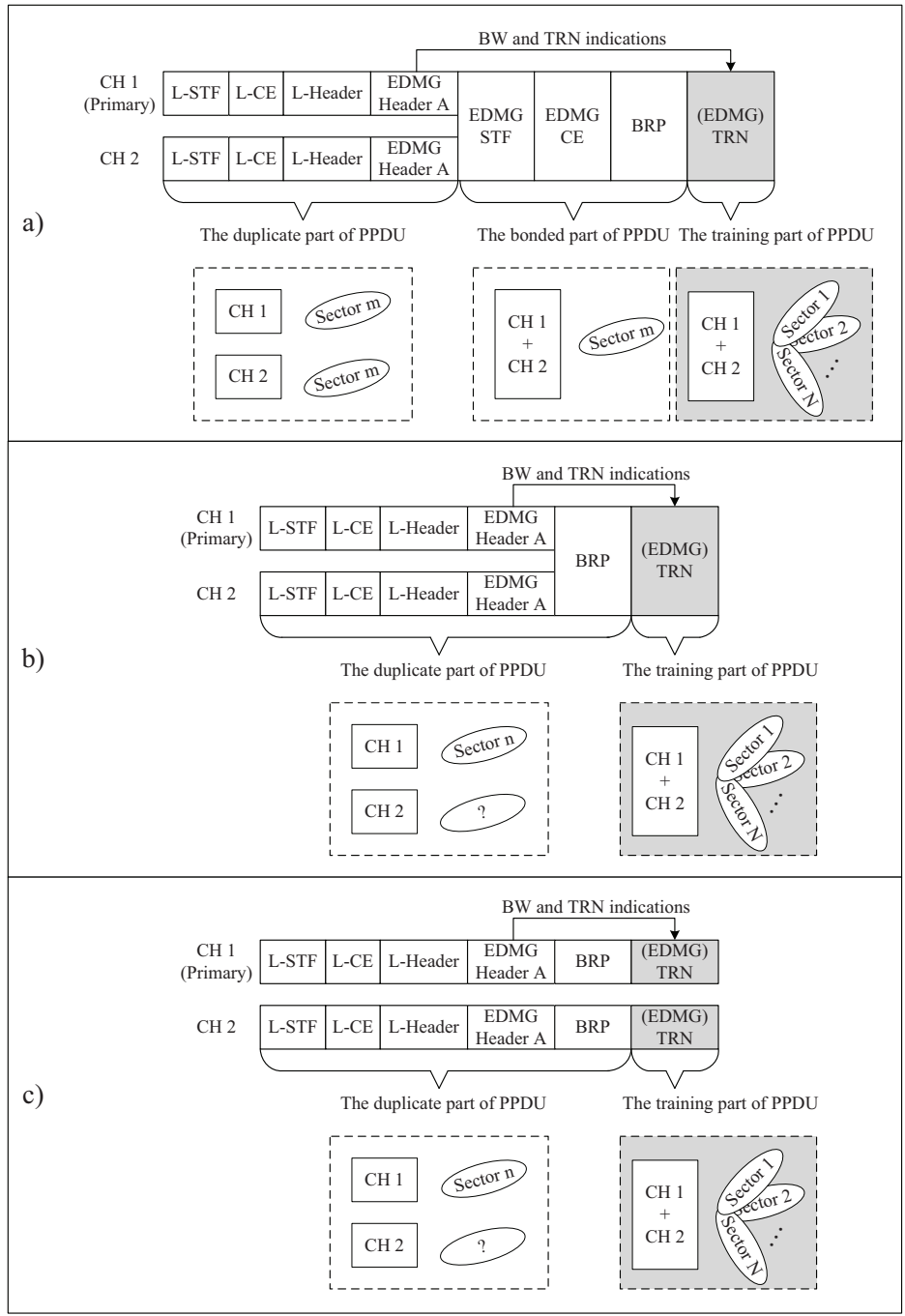}}
    \caption{ a) Existing BRP PPDU format, beamforming on a bonded channel is trained. b) Proposed BRP PPDU format 1, beamforming on a bonded channel is not trained. c) Proposed BRP PPDU format 2, beamforming on an aggregated channel is not trained.}
  \end{center}
\end{figure*}

\subsection{Hybrid beamforming}

\begin{figure*}[!htbp]
  \begin{center}
    \scalebox{0.85}[0.85]{\includegraphics{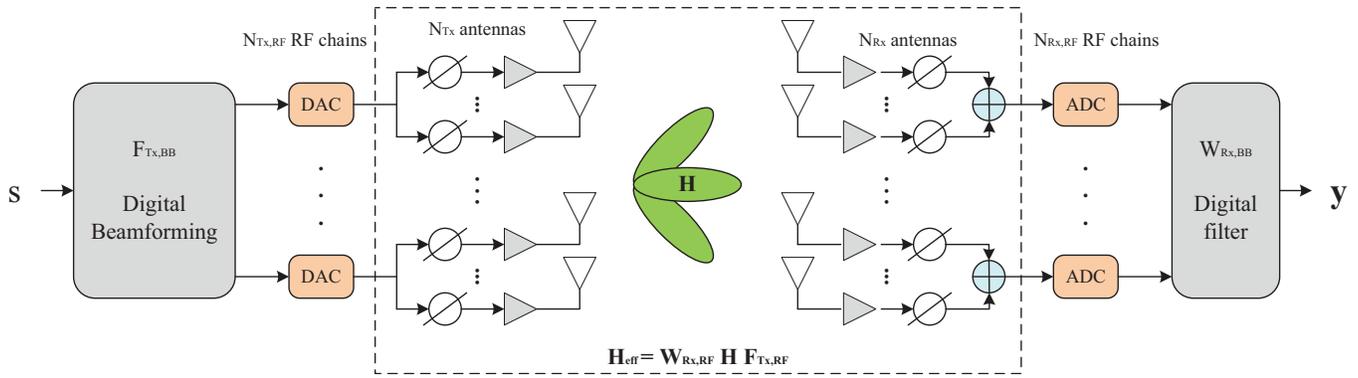}}
    \caption{Hybrid beamforming architecture.}
  \end{center}
\end{figure*}

Hybrid beamforming is a combination of digital beamforming and analog beamforming recommended for 802.11ay \cite{ref58}. Since there is hardly consensus reached on hybrid beamforming for the 802.11ay, we only present the architecture and the basic procedures for hybrid beamforming with effective channel information. The architecture of hybrid beamforming is shown in Fig. 22. Analog beamforming is responsible for increasing the link gain and digital beamforming is responsible for increasing the spatial multiplexing gain. The received signal $y$ = ${W}_{RxBB}$${W}_{RxRF}$$H$${F}_{TxRF}$${F}_{TxBB}$$s$ + $n$ = ${W}_{RxBB}$${H}_{eff}$${F}_{TxBB}$$s$ + $n$, where ${W}_{RxBB}$ is the baseband beamformer of the receiver, ${W}_{RxRF}$ is the analog beamformer of the receiver, $H$ is the channel matrix (full channel state information (CSI)), ${F}_{TxRF}$ is the analog beamformer of the transmitter, ${F}_{TxBB}$ is the baseband beamformer of the transmitter, $s$ is the transmitted signal, $n$ is the noise, and ${H}_{eff}$ = ${W}_{RxRF}$$H$${F}_{TxRF}$ is the effective channel. It is difficult to estimate full CSI in mmWave systems, because we should obtain enough Signal-Noise-Ratio (SNR) by performing time-consuming BFT processes. How to design an efficient feedback scheme of the large size full CSI is another problem. Therefore, Kim \emph{et al.} \cite{ref59} proposed a scheme to estimate the effective channel matrix instead of the full channel matrix once analog beams are determined and the transmitter needs to know the effective channel matrix (or simply CSI) to obtain beamforming gain. The procedure for hybrid beamforming with effective channel information proposed by \cite{ref59} contains three steps. Firstly, it shall obtain the analog beamforming/filter by performing the BFT (e.g., SLS and BRP) without considering the effect of digital beamforming/filter (i.e., ${W}_{BB}$ = I and ${F}_{BB}$ = I). Secondly, the receiving STA (the receiver) estimates the effective channel (${H}_{eff}$) through the pilot sequences transmitted by the PCP/AP (the transmitter) and then feeds the effective channel information back to the PCP/AP. Thirdly, the PCP/AP designs its digital beamforming based on the effective channel information, while STA obtains its digital beamforming based on the estimated effective channel.

A poor beamformed link or an inefficient BFT procedure will decrease the system throughput and influence the QoE. Therefore, from Section V-A to Section V-F, there are many effective mechanisms proposed to improve BFT performance. In addition, Mubarak \emph{et al.} \cite{ref60} proposed a BFT, discovery and association mechanism to reduce the complexity of the mmWave link establishment by using Wi-Fi positioning technology. The main idea is similar to the ILA mechanism proposed in \cite{ref53} and Binary Search Beamforming \& Linear Search Beamforming mechanisms proposed in \cite{ref61}. They all use location information to reduce the complexity of BFT. Another way to reduce BFT time and provide a robust BFT experience in the non-LOS (NLOS) mmWave environment proposed in \cite{ref62} is beam coding. By assigning each beam a unique signature code, multiple beams can steer at their angles simultaneously in the training frame, which will benefit the BFT efficiency. Actually, there are still some worthy beamforming mechanisms under discussions in the development of the 802.11ay, such as SU-MIMO beamforming and DL MU-MIMO beamforming, which will be introduced in Section VI.

\subsection{Beamforming for asymmetric links}
In the 802.11ad and the 802.11ay, the use of quasi-omni antenna configuration may result in asymmetric links. The characteristic of an asymmetric link is that an STA can receive frames from the peer STA, but the peer STA cannot receive frames from it. Therefore, a mechanism, named beamforming for asymmetric links, was proposed in the 802.11ay \cite{ref4} to cope with this issue. Firstly, a PCP/AP appends TRN-R fields to DMG Beacon frames to perform BTI and after STAs receiving the DMG Beacon frames, they perform receive beamforming using these TRN-R fields as described before. Secondly, each STA transmits an SSW or Short SSW frame in the directional mode through the best sector trained by TRN-R (antenna reciprocity should be assumed in this procedure). At the same time, the PCP/AP listens for these SSW or Short SSW frames and the sector listening order is the same as the DMG Beacon frames transmitted during the BTI. Thirdly, the STAs go to the directional receive mode after finishing the SSW or Short SSW frames transmissions. Then, the PCP/AP transmits a Sector ACK frame in each sector where an SSW or Short SSW frame is received. Finally, the PCP/AP can allocate directional transmissions with the STAs and all the frame exchanges between the PCP/AP and the STAs will use the established beamformed links.

For example, as shown in Fig. 23, the links from the PCP/AP to STAs have been established through the TRN-R fields in the last BTI phase. EDMG STA 1 chooses Sector ID = 0 to transmit the Short SSW frame in the direction trained through the TRN-R. EDMG STA 2 chooses Sector ID = 1 for the same purpose. At the same time, the PCP/AP listens for the Short SSW frames in the same order of transmitting the DMG Beacon frames. Then, the PCP/AP transmits the Sector ACK frames through the same sectors where the SSW or Short SSW frames are received and the EDMG STAs listen directionally with their beamforming links. After finishing these processes, beamforming for asymmetric links is said to be done. The details can be found in \cite{ref4}.

\begin{figure*}[!htbp]
  \begin{center}
    \scalebox{0.65}[0.65]{\includegraphics{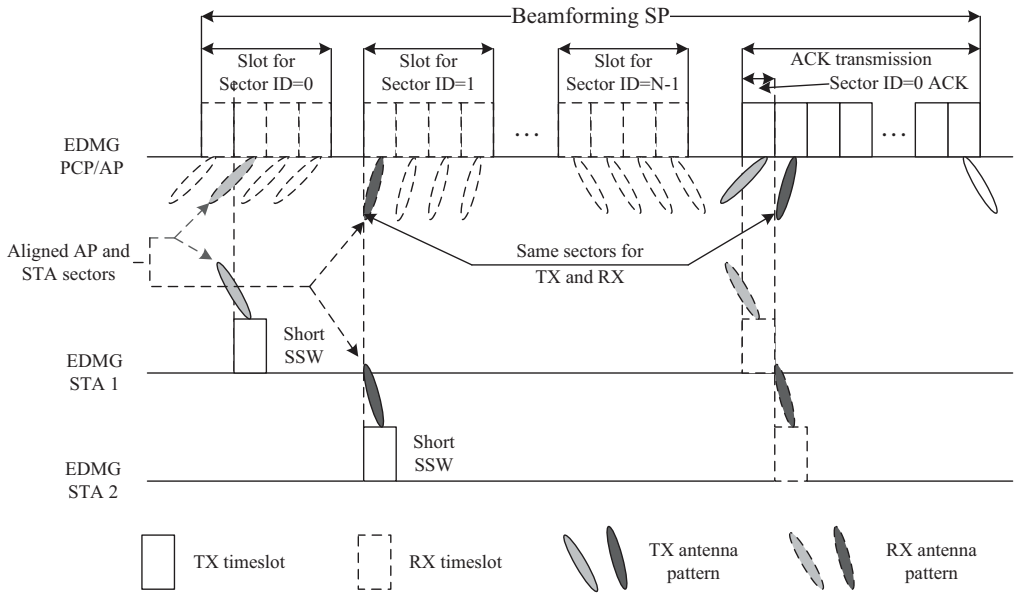}}
    \caption{Beamforming for asymmetric links.}
  \end{center}
\end{figure*}

\subsection{Beam tracking}

\begin{figure}[!htbp]
  \begin{center}
    \scalebox{0.42}[0.42]{\includegraphics{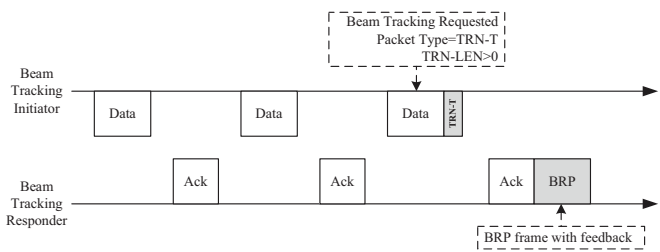}}
    \caption{An example of beam tracking with an initiator requesting TRN-T.}
  \end{center}
\end{figure}

\begin{figure*}[bp]
  \begin{center}
    \scalebox{0.65}[0.65]{\includegraphics{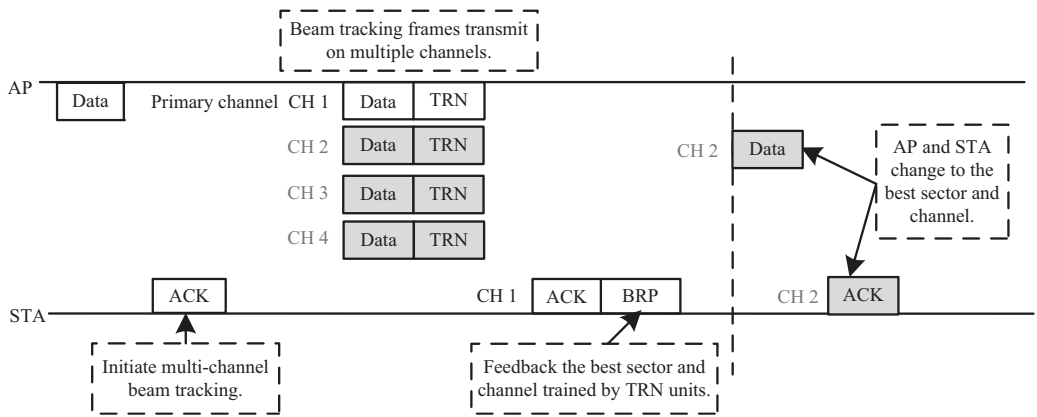}}
    \caption{Multi-channel beam tracking.}
  \end{center}
\end{figure*}

In the 802.11ad, if the quality of a beamformed link is not suitable for data transmissions any more (generally the SNR is lower than a specified threshold but the link has not been blocked), the beam tracking initiator can append TRN units to data frames to perform beam tracking. As shown in Fig. 24, performing beam tracking is more efficient and flexible compared with redoing BFT when the beamformed link's quality decreases. The last `Ack + BRP' frame will feed back the beam tracking results trained by TRN-T fields. Thus, the beam tracking initiator can select an optimal beam to transmit the remaining data frames. Two types of beam tracking for 802.11ay were proposed in \cite{ref63}, one is similar to the beam tracking method defined in 802.11ad, namely analog beam tracking (ABT), and the other is digital baseband channel tracking (DBC). The ABT (a TRN units based training) method can be used to cope with STA rotation, movement or blockage by tracking the changes in analog beams. The DBC (a TRN units or CEF based training) method can be used to deal with beam blockage when using hybrid beamforming by tracking the changes in the baseband channel for a fixed set of analog beams.

Since 802.11ay supports multi-channel operation, STAs can be allocated with different channels during the DTI through the DMG Beacon frame in the BTI or the announcement frame in the ATI. Suppose an STA is allocated with multiple channels, if the operating channel is blocked and an available beam cannot be found by performing beam tracking procedure as depicted in Fig. 24, the STA can perform beam tracking on other channels allocated to it. We propose a multi-channel beam tracking mechanism, as shown in Fig. 25, in which an STA can initiate the multi-channel beam tracking procedure by transmitting an Ack frame with `multi-channel beam tracking requested', then the AP transmits the `Data + TRN' frames on the channels allocated to it to perform multi-channel beam tracking. Next, the STA feeds back the results on primary channel (channel 1) and then changes to channel 2 that is the best channel obtained from multi-channel beam tracking procedure. Finally, the AP transmits the remaining data frames on channel 2 with the best sector as indicated in the BRP frame appended to the Ack frame transmitted by the STA.

\subsection{Beamformed link robustness}
Due to the serious path-loss and difficulties in keeping beams aligned, 60 GHz mmWave frequency bands are very sensitive to blockage and mobility. To cope with this issue, the 802.11ad and the 802.11ay provide some effective methods such as Fast Session Transfer (FST), beam tracking and relay operation. The FST is a method that allows communications in 60 GHz band fallback to 2.4/5 GHz, which is of great importance in providing seamless multi-Gbps coverage. In fact, cooperative joint network design over low frequency band (e.g., 2.4/5 GHz) and high frequency band (e.g., 60 GHz) is a promising trend in future WLANs and 5G cellular networks, which demands further research.

The second method is beam tracking. The main idea to fight against blockage through beam tracking is to find another pair of available beams to continue transmission between the two STAs. We have introduced the beam tracking procedure defined in 802.11ad and proposed a multi-channel beam tracking mechanism in Section V-G. A double-link beam tracking mechanism to fight against human blockage and device mobility was proposed in \cite{ref64}, which includes a search strategy of transmissions and alternative links and adopts in-packet double-link tracking and switching. As shown in Fig. 26, Gao \emph{et al.} suggested to append an alternative-link beam tracking field at the end of data frame (or other frames) \cite{ref64}. Thus, the original TRN units can train the transmission link and the TRN units in alternative-link beam tracking can train the alternative link. It is highly unlikely for both the transmission link and alternative link to be blocked simultaneously. Therefore, compared with beam tracking procedures in 802.11ad, the double-link beam tracking can significantly reduce the outage probability and improve the network throughput.

\begin{figure}[!htbp]
  \begin{center}
    \scalebox{0.42}[0.42]{\includegraphics{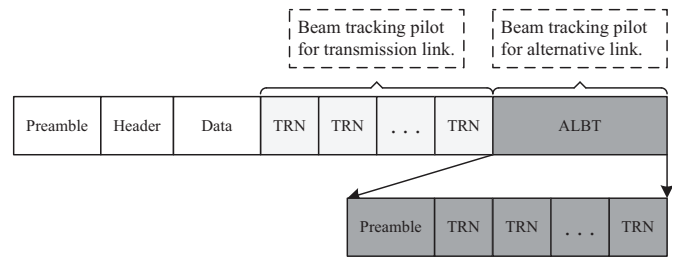}}
    \caption{Frame design for double-link beam tracking.}
  \end{center}
\end{figure}

Xue \emph{et al.} in \cite{ref40} proposed a cooperative beam tracking mechanism in beamspace SU-MIMO mmWave wireless communications. In the beamspace SU-MIMO (an mmWave communication mode in which multiple beams can be supported at both transmitter and receiver), two STAs communicates through multiple concurrent beam pairs. If one of the concurrent transmission links is blocked, we can restore the blocked link with the help of the unblocked links, since beam tracking and beam switching signaling can be transmitted through the unblocked links. The double-link beam tracking mechanism and cooperative beam tracking mechanism will be of great importance in future mmWave communications to resolve blockage problems.

The third method is relaying. Relay links can be considered as backup for LOS transmissions, and can be utilized when the direct link fails in the 802.11ad \cite{ref65}. However, there are several deficiencies in the 802.11ad relay operations. For example, the channel time for Source-Relay and Relay-Destination data transfer is fixed (and retransmissions would be inefficient), which lacks support for multi-hop relaying and does not support efficient bi-directional data transfer. Unfortunately, there do not seem to have any significant improvements as yet for the 802.11ay. Here, we review some enhanced relay operations for mmWave communication systems (especially in the 802.11ad) proposed in the current literature. Zheng \emph{et al.} \cite{ref66} considered the problem of robust relay placement in 60 GHz mmWave WPANs with directional antenna, and formulate the robust minimum relay placement problem and robust maximum utility relay placement, for better connectivity and robustness to cope with link blockage. Although decode-and-forward relays can extend the limited coverage and deal with blockage for mmWave communications, it decreases the effective throughput (defined as the successfully transmitted bits over the duration between two available sequential time slots), which will affect the data-greedy applications. Owing to the good space isolation brought by significant path-loss, Lan \emph{et al.} \cite{ref67} proposed a deflection routing scheme, in which time slots can be shared between direct path and relay path to improve effective throughput for mmWave WPAN networks. A sub-exhaustive search based best fit deflection routing algorithm was proposed to find a relay path which can maximize the effective throughput while bringing in the least interference. Similarly, in order to cope with the blockage problem in mmWave networks, Niu \emph{et al.} \cite{ref26} also addressed relaying and proposed a frame based blockage robust and efficient directional medium access control protocol. It jointly optimizes the relay selection and the spatial reuse to improve effective throughput. It has been shown that it provides better delay, throughput and fairness performance than other existing protocols.

\section{SU-MIMO and MU-MIMO Beamforming}

\subsection{General Overview}
The 802.11ay supports both SU-MIMO and MU-MIMO technologies, which can provide robust data transmissions and high data rates by using multiple directional beams simultaneously. However, SU-MIMO beamforming and MU-MIMO beamforming are different from that of DMG beamforming described in Section V. Recently, SU-MIMO and MU-MIMO beamforming technologies are gaining intensive attention from both academia and industries \cite{ref3}, \cite{ref68, ref69, ref70, ref71, ref72, ref73, ref74}, \cite{ref76, ref77, ref78, ref79}. The details are presented in the following few subsections.

\subsection{SU-MIMO beamforming}

\begin{figure*}[bp]
  \begin{center}
    \scalebox{0.65}[0.65]{\includegraphics{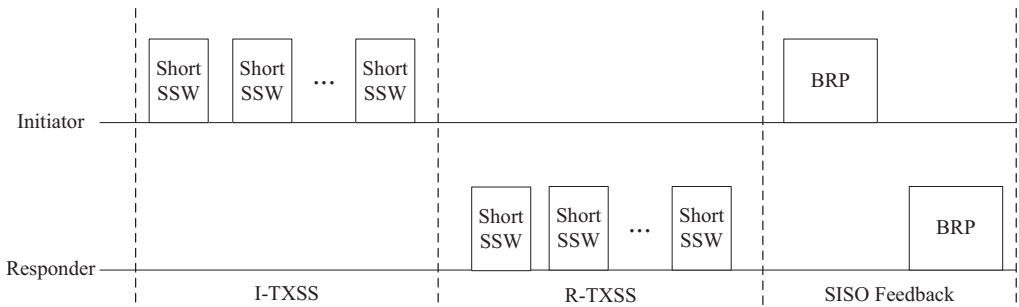}}
    \caption{SU-MIMO SISO phase beamforming procedure.}
  \end{center}
\end{figure*}

\begin{figure*}[bp]
  \begin{center}
    \scalebox{0.75}[0.75]{\includegraphics{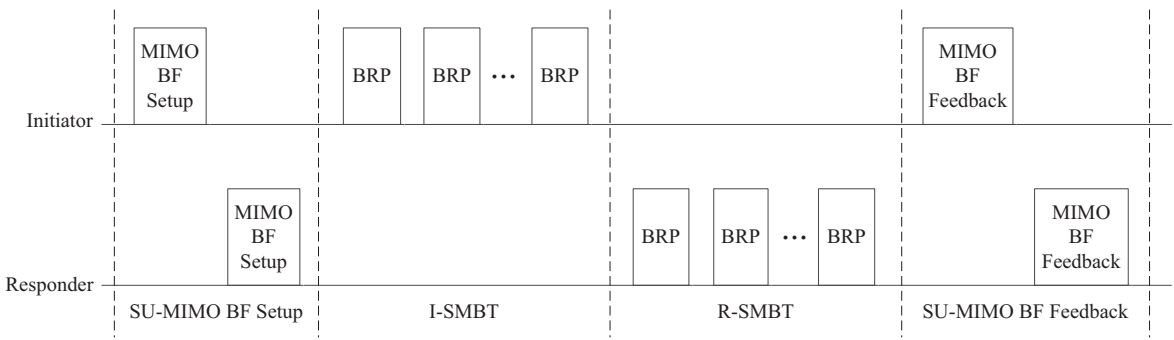}}
    \caption{SU-MIMO MIMO phase beamforming procedure.}
  \end{center}
\end{figure*}

The SU-MIMO beamforming protocol supports the BFT for transmissions and receptions of multiple spatial streams between an SU-MIMO capable initiator and an SU-MIMO capable responder \cite{ref4}, \cite{ref68}, \cite{ref69}. The SU-MIMO beamforming protocol enables the configuration of TX antenna settings and the corresponding RX antenna settings for simultaneous transmissions of multiple spatial streams from the initiator to the responder or vice versa. The SU-MIMO beamforming protocol shall be performed in the DTI phase of the BI and shall not be initiated unless the initiator and the responder have a valid SISO beamformed link between them. After SU-MIMO BFT procedure is finished, both the initiator and the responder obtain some well-trained transmission and reception antenna settings. There are two phases in SU-MIMO BFT, namely, SU-MIMO SISO phase and SU-MIMO MIMO phase, which are shown in Fig. 27 and Fig. 28, respectively.

SU-MIMO SISO phase comprises three subphases, including I-TXSS, R-TXSS and SISO feedback as shown in Fig. 27, where the first two subphases are optional whereas the last one is mandatory. The goal of the SISO phase is to enable the initiator to collect feedback of the last I-TXSS from the responder and also enables the responder to collect feedback of the last R-TXSS from the initiator. The initiator may perform the I-TXSS subphase to start the SISO phase, during which the initiator transmits Short SSW frames through directional scanning to accomplish transmission sectors training so as to get the best TX sector of the initiator. If the I-TXSS subphase was present in the SISO phase, the responder shall initiate the R-TXSS subphase following the completion of the I-TXSS subphase. Otherwise, the responder shall not initiate the R-TXSS subphase. During the R-TXSS subphase, the responder transmits the Short SSW frames through directional scanning to accomplish transmission sectors training so as to get the best TX sectors of the responder. During the SISO feedback subphase, the initiator and the responder exchange the BRP frames alternatively. These BRP frames contain channel measurement feedback element which includes channel measurement information obtained in the previous TXSS by the counterpart when scanning each sector. Through the aforementioned three subphases, the initiator gets the feedback information obtained in I-TXSS from the responder, while the responder gets the feedback information obtained in R-TXSS from the initiator.

\begin{figure*}[bp]
  \begin{center}
    \scalebox{0.7}[0.7]{\includegraphics{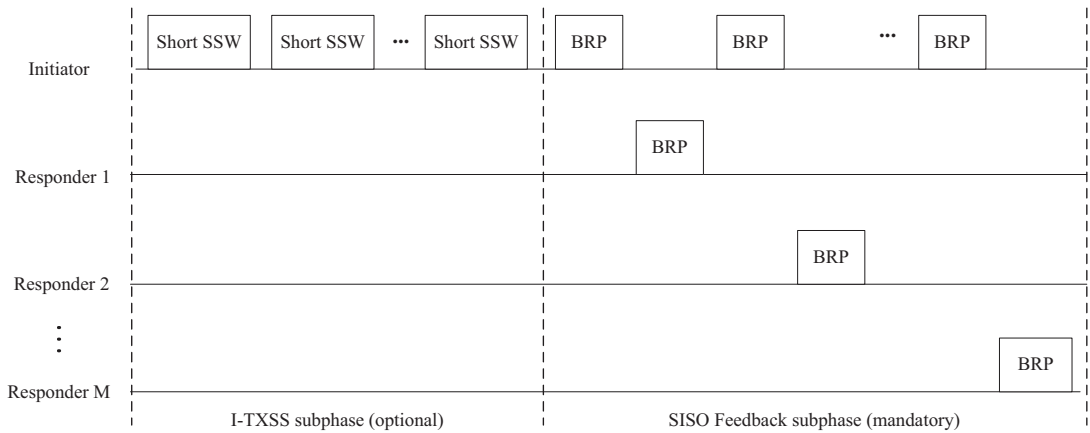}}
    \caption{MU-MIMO SISO phase beamforming procedure.}
  \end{center}
\end{figure*}

\begin{figure*}[bp]
  \begin{center}
    \scalebox{0.8}[0.8]{\includegraphics{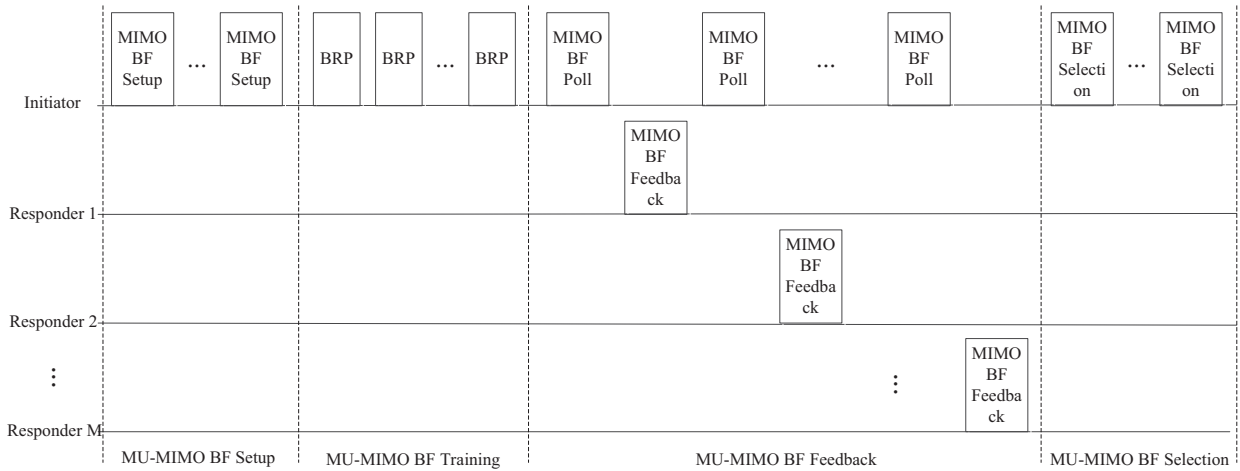}}
    \caption{MU-MIMO MIMO phase beamforming procedure.}
  \end{center}
\end{figure*}

On the other hand, SU-MIMO MIMO phase consists of four subphases, namely, SU-MIMO BF setup, initiator SU-MIMO BF training (I-SMBT), responder SU-MIMO BF training (R-SMBT), and SU-MIMO BF feedback as shown in Fig. 28. The initiator shall start with the MIMO phase after an MBIFS following the end of the SISO phase. SU-MIMO MIMO phase is mainly used to enable the training of TX and RX sectors and DMG antennas to determine the best combinations of TX and RX sectors and DMG antennas for SU-MIMO operations. During the mandatory SU-MIMO BF Setup subphase, the initiator and responder shall transmit a MIMO BF setup frame to inform the counterpart the following information:

(1) a unique dialog token for identifying SU-MIMO BF training,

(2) the number of transmit sector combinations requested for the initiator link and the responder link,

(3) whether time domain channel response is requested as part of SU-MIMO BF feedback,

(4) the number of TRN subfields requested for receive AWV training in the following responder and initiator SMBT subphase.

During the I-SMBT subphase, the initiator transmits the EDMG BRP-RX/TX frames to accomplish the training of multiple TX and RX sectors for SU-MIMO operation of the initiator link. The responder shall perform the R-SMBT subphase following the completion of the I-SMBT subphase, during which the responder transmits the EDMG BRP-RX/TX frames to accomplish the training of multiple TX and RX sectors for SU-MIMO operations of the responder link. The SU-MIMO BF feedback subphase generally contains four cases, and more details are given in \cite{ref4}.

\subsection{MU-MIMO beamforming}

MU-MIMO BFT is also completed in the DTI phase of the BI, which is employed to establish a DMG antenna configuration for an initiator transmitting the EDMG MU PPDU to the responders that have MU-MIMO capability, so as to make the mutual interference among multiple data streams carried by the MU PPDU to the lowest \cite{ref4}, \cite{ref70}, \cite{ref71}. It contains two consecutive phases similar to the SU-MIMO BFT, namely MU-MIMO SISO phase and MU-MIMO MIMO phase. The MU-MIMO SISO phase includes two subphases: the optional I-TXSS subphase and the mandatory SISO Feedback subphase as shown in Fig. 29. During the MU-MIMO SISO phase, the initiator and the responders who intend to participate in the MU group, shall train the TX and RX antennas and sectors and collect the feedback information.

\begin{figure*}[!htbp]
  \begin{center}
    \scalebox{0.83}[0.83]{\includegraphics{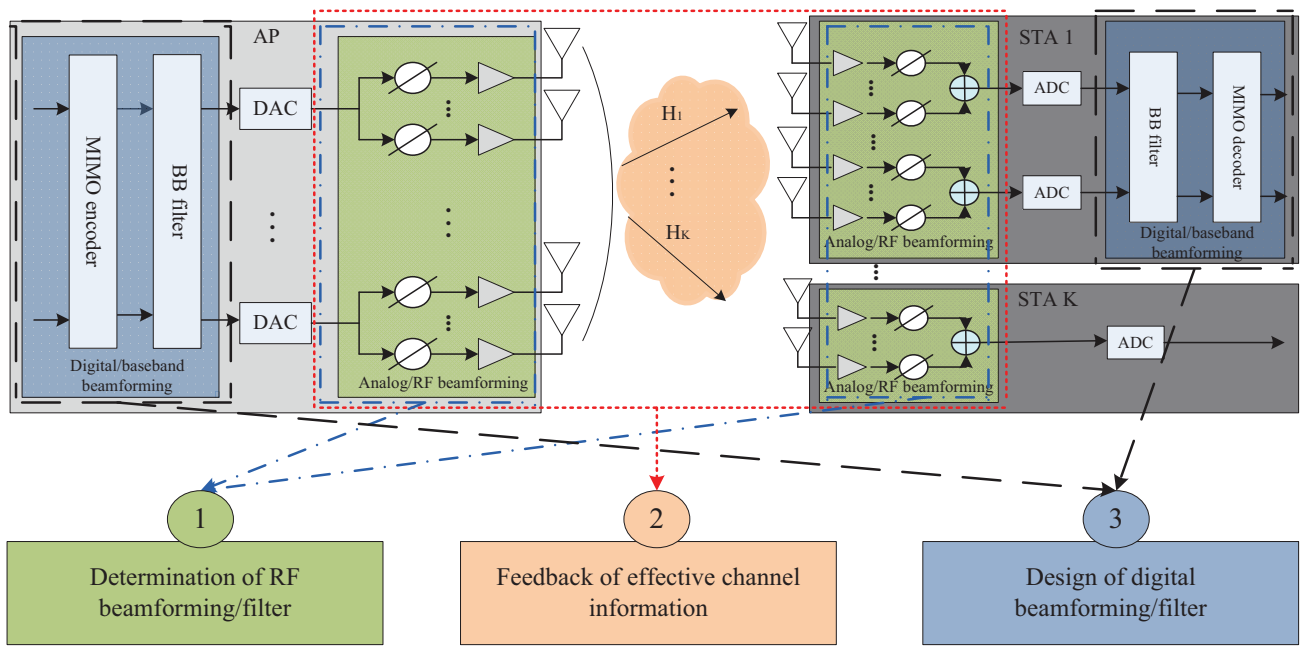}}
    \caption{Procedure of designing MU-MIMO hybrid beamforming.}
  \end{center}
\end{figure*}

During the I-TXSS subphase, the initiator transmits the Short SSW frames to train the TX and RX sectors between the initiator and the responders who intend to join the MU group. During the SISO Feedback subphase, the initiator will transmit a BRP frame to poll every responder who intends to join MU group. In this way, the initiator can obtain the sector list of each TX DMG antenna and the associated quality indication between the initiator and each responder. After the responder receives a BRP frame, it shall respond with a corresponding BRP frame, which includes the sectors and its associated quality indication of each initiator's TX DMG antenna. Once the SISO phase is completed, the initiator collects the feedback information of the TX and RX antennas and sectors on initiator link between the initiator and the responders who intend to join the MU group.

As shown in Fig. 30, the MIMO phase of MU-MIMO BFT consists of four subphases, namely, an MU-MIMO BF setup subphase, an MU-MIMO BF training subphase, an MU-MIMO BF Feedback subphase, and an MU-MIMO BF selection subphase. During the MU-MIMO BF setup subphase, the initiator shall transmit one or more BF setup frames to each responder in the MU group, which indicates the following information:

(1) the EDMG group ID of the MU group,

(2) each remaining responder in the Group User Mask field,

(3) a unique dialog token in the Dialog Token field for identifying MU-MIMO BF training,

(4) whether time domain channel response is requested as part of MU-MIMO BF feedback.

To reduce the MU-MIMO BFT time, the initiator may select a TX sector subset for each DMG antenna according to the feedback from the responders. It is worth pointing out that in the MU-MIMO BF setup subphase, the initiator may exclude some responders from the following MU-MIMO BFT subphase and the MU-MIMO BF poll subphase if multi-user interference due to MU-MIMO transmissions they suffer is expected to be negligible according to the feedback of the SISO phase. During the MU-MIMO BFT subphase, the initiator shall transmit one or more EDMG BRP-RX/TX frames to the remaining responders in the MU group to train one or more TX sectors. While during the MU-MIMO BF Feedback subphase, the initiator shall transmit a MIMO BF Feedback Poll frame carrying the dialog token that identifies the MU-MIMO training to poll each intended responder to collect MU-MIMO BF feedback from the preceding MU-MIMO BFT subphase. Once the responder receives the MIMO BF Poll frame, the responder shall respond a MIMO BF Feedback frame which carries the list of received initiator's TX DMG antennas/sectors, each with its corresponding responder's RX DMG antenna/sector and the associated quality indicated. In the MU-MIMO BF selection subphase, the initiator shall transmit a BF selection frame to each responder in the MU group containing the dialog token identifying the MU-MIMO training, one or multiple sets of the MU transmission configurations, and the intended recipient STAs for each MU transmission configuration. After the above mentioned subphases completed, multiple responders in the MU group and their corresponding MU transmission configurations can be selected.

However, there is a significant challenge for MU-MIMO in mmWave WLANs. With the increase of the number of STAs and the size of a codebook, the overhead required to determine the RF/digital beams exponentially increases. Therefore, a low complexity and low overhead hybrid beamforming algorithm is required. To this end, Kim \emph{et al.} \cite{ref72} discussed the need for MU-MIMO and the parameters that should be determined to support MU-MIMO in the 802.11ay, and consider the hybrid beamforming to support DL MU-MIMO transmission in 802.11ay. A three-stage approach to design an MU-MIMO hybrid beamforming was developed as shown in Fig. 31. More specifically, in the first stage, RF beamforming/filter for each user is selected based on the BFT without considering the effect of digital beamforming/filter. In the second stage, the AP transmits the pilot sequences, while each STA estimates and quantizes its effective channel and feeds back the index of the quantized channel to the AP. The AP calculates the digital beamforming/filter based on the quantized channels and forwards the digital filters to STAs in the third stage. Considering the BFT for hybrid MIMO in the 802.11ay may be expensive as the number of possible beam combinations grows exponentially with the number of antenna arrays, a low complexity BFT method for the 802.11ay by using evolutionary BFT and \emph{K}-best BFT was proposed in \cite{ref73}, where the simulation results show that there is a significant reduction in complexity and a negligible loss in the MIMO capacity compared to exhaustive search. It makes hybrid MU-MIMO more practical to implement.

Training with Tx/Rx beam sectorization is a current training method for DL MU-MIMO in the 802.11ay, which is transparent to any UL-DL channel differences. However, there is a challenge that this method is time consuming (as much training as the number of sectors) and the resulting beam accuracy is low (higher accuracy will demand more training overhead). To tackle these issues, an UL training method for DL MU-MIMO was proposed in \cite{ref74} as shown in Fig. 32, which has less training overhead and superior beam accuracy, but needs possibly more precise calibration due to the lack of UL/DL channel reciprocity. In this approach, the AP first transmits an UL training trigger frame to a specific group of STAs, then each STA in the group responds with a training packet. After that, the AP estimates the received channel (or any transformation of the channel), and then uses the estimated channel for DL transmissions. It is assumed that the calibration has been performed before conducting the UL training procedure with the method presented in \cite{ref75}. In practical implementations, DL MU-MIMO throughput might be degraded due to CSI feedback overhead. However, when the AP can form multiple narrow beams, it enables DL MU-MIMO without CSI feedback theoretically. When each STA reports the SNR \& Channel measurement with MCS4 in the 802.11ad, the CSI feedback takes about 3\textmu s $\thicksim$ 4\textmu s. Thus, the total duration of CSI feedback from all STAs in the MU-MIMO operations is estimated to be several \textmu s. Therefore, the DL MU-MIMO operation that causes little interference between STAs without CSI feedback is preferred in \cite{ref76}.

\begin{figure}[!htbp]
  \begin{center}
    \scalebox{0.43}[0.43]{\includegraphics{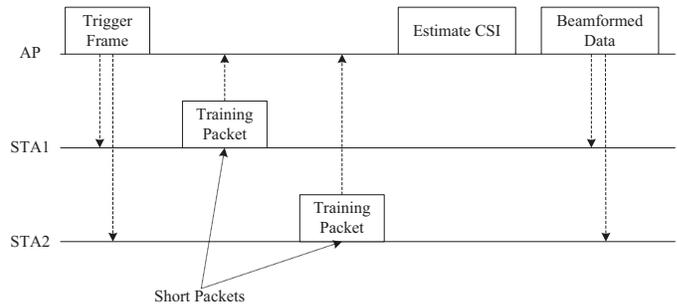}}
    \caption{UL training procedure for DL MU-MIMO.}
  \end{center}
\end{figure}

When an AP forms wider beams for DL MU-MIMO operations, large interference may occur among beams, and thus CSI is needed to mitigate interference. However, when an AP forms very narrow beams for DL MU-MIMO operations, large interference may not occur among beams, which means CSI feedback is not needed theoretically. In some cases, the signal power received by an STA might be insignificant unless the best sector is used, where the interference at the STA in MU transmissions might be small even without MU-MIMO BFT. Inspired by this idea, Fujio \emph{et al.} \cite{ref77} proposed a method to enable an initiator to exclude some responders from the MU-MIMO BFT and the MU-MIMO BF polls subphases in the MIMO phase of MU-MIMO beamforming as shown in Fig. 33. Specifically, the initiator collects feedbacks such as Sector IDs and their corresponding SNRs from the responders in the SISO phase. Based on these feedbacks, the initiator can estimate whether multiuser interference in MU transmissions is small at each of the responders. If the interference at some of the responders is expected to be small, the initiator may exclude them from the subphases. In \cite{ref78}, Li \emph{et al.} generalized the joint spatial division multiplexing scheme to support non-orthogonal virtual sectorization and hybrid analog/digital structures at both BS and UEs, where the analog precoders and combiners are based on the second order channel statistics, and the digital combiners are based on both intra and inter-group instant effective channels, while the digital precoders are only based on intra-group channels at the BS and the second order channel statistics. In \cite{ref79}, Qiao \emph{et al.} studied concurrent beamforming issue for achieving high capacity in indoor mmWave networks, where the general concurrent beamforming issue was formulated as an optimization problem to maximize the sum rates of concurrent transmissions by considering the mutual interference. To reduce the complexity of beamforming and the total setup time, the concurrent beamforming is decomposed into multiple single-link beamforming, and an iterative searching algorithm was proposed to quickly achieve the suboptimal transmission/ reception beam sets. A codebook-based beamforming protocol at MAC is then introduced in a distributed manner to determine the beam sets. Due to the requirement of establishing several independent links, multi-device multi-path BFT becomes rather time-consuming in 60 GHz mmWave communications. In fact, the aforementioned methods for DL MU-MIMO are feasible in theory, but they may too complex to implement in current mmWave WLANs.

\begin{figure}[!htbp]
  \begin{center}
    \scalebox{0.43}[0.43]{\includegraphics{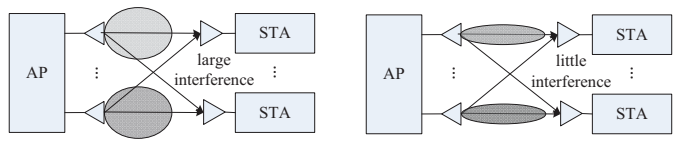}}
    \caption{DL MU-MIMO w/o CSI.}
  \end{center}
\end{figure}

\section{Open Research Issues and Future Work}
In the near future, WLANs will be more advanced with new technologies incorporated as we have discussed in this paper. Owing to the high-efficiency and flexible characteristics of cellular systems \cite{ref80}, the evolution of WLANs will inherit many features of the cellular systems. In this section, we will discuss some open issues and research directions for future generation WLAN systems, including inter-AP cooperation, HF and LF cooperation, dual connectivity, power management, UL MU-MIMO, group beamforming, and Cloud-RAN based WLANs.

\subsection{Inter-AP cooperation}
In cellular systems, control signaling can be exchanged via X2 interface among evolved NodeBs (eNodeBs) \cite{ref81}. Thus, handoff among eNodeBs will be easier than before. For the future WLANs, inter-AP cooperation will be an important feature. If an STA suffers from poor link quality, the serving AP can inform the other candidate APs nearby to prepare necessary information for AP handoff through inter-AP cooperation. Once an STA's blockage occurs with a serving AP due to obstacles, the APs that cooperate with the serving AP can establish other links with the STA. Therefore, fast AP handoff, fast link recovery, joint transmissions from two or more Aps, and other benefits can be realized with inter-AP cooperation. Thus, how to design an efficient inter-AP cooperation mechanism will be of great importance for the future WLANs.

\subsection{High frequency and low frequency cooperation}
Taking the heterogeneous cellular architecture as an example, there is one macro cell operating in low frequency that covers a broad area, while there are many small cells co-locating under the coverage of a macro cell \cite{ref82}, \cite{ref83}. Each small cell operates in high frequency (e.g., mmWave frequency band) and only covers a limited area. A UE is connected to the macro cell and one or more small cells simultaneously. Control plane and part of the user plane traffics can be transmitted through the macro cell to guarantee reliable communications, while most of the user plane traffics are transmitted through small cells to provide high data rates \cite{ref82}. This kind of control plane/data plane decoupling architecture (C/U decoupling) can take full use of both the HF and LF \cite{ref84}. For future generation WLANs, HF and LF cooperation can be an important feature, since 2.4 GHz, 5 GHz and 60 GHz have already used in the existing WLANs. The FST mechanism defined in \cite{ref2} allows the traffics in 60 GHz transferring to 2.4 GHz/5 GHz in case of the blockage in 60 GHz. HF and LF cooperation can provide better mobility management. When an STA moves out of the beam coverage of its serving AP, HF link failure will occur between STA and the serving AP. However, the LF link may still keep connection with the AP and the connection will not be interrupted. In addition, LF can assist HF to perform BFT, beam tracking, etc. We believe that, HF and LF cooperation, which can bring many improvements for future WLANs, needs further consideration.

\subsection{Dual connectivity}
Dual connectivity will be widely used in the aforementioned heterogeneous cellular architecture \cite{ref85}. HF \& LF cooperation and inter-AP cooperation will all adopt dual connectivity. The main difference is that, an STA connects both LF and HF in HF and LF cooperation, but it may connect two HF APs or two LF APs or one HF AP and one LF AP in dual connectivity. With dual connectivity, Semiari \emph{et al.} proposed a novel MAC to dynamically manage the WLAN traffic over the unlicensed mmWave and sub-6 GHz bands \cite{ref86}. Therefore, the saturation throughput can be maximized and the delay can be minimized. In addition, a novel dual-mode scheduling framework was also proposed by Semiari \emph{et al.} to increase the number of satisfied user applications while curtailing the QoS violations and enhancing the overall users' QoE by jointly performing user applications selection and scheduling over LF and HF \cite{ref87}. If an STA connects with two APs by dual connectivity, when it moves out of the coverage of one AP, the other AP can provide communication service without disruption. The STA may find a new AP and establish a link with the new AP before ``the other link'' is interrupted. Thus, soft handoff and better mobility management can be realized through dual connectivity. There are still many benefits to be explored for the future WLANs with dual connectivity.

\subsection{Power management}
For devices equipped with large antenna arrays to form a narrow beam for directional communications, power consumption in mmWave communications is a serious issue \cite{ref27}. Especially when dual connectivity and HF \& LF cooperation are adopted, multiple transceivers should keep in active mode to maintain multiple links. It will be a great challenge for mobile devices equipped with batteries. To avoid unnecessarily staying in active mode, the related STAs or transceivers should go to the sleep mode immediately. How to design effective power save mechanisms with HF and LF cooperation, inter-AP cooperation, and dual connectivity is of paramount importance, which forms a fruitful research direction. In addition, intelligent and efficient power save mechanisms should be carefully designed to cope with power consumption across multiple design dimensions.

\subsection{Uplink multi-user MIMO}
UL MU-MIMO has been accepted in the 802.11ax standard \cite{ref88}, \cite{ref89}. For the future WLANs, UL MU-MIMO is still a hot topic, especially in mmWave WLAN systems. Since directional transmissions in mmWave communications will incur less interference, adopting beamspace UL MU-MIMO in mmWave WLAN systems will be feasible. According to the widely adopted UL MU-MIMO feature in cellular systems, how to deal with the timing offset and the frequency offsets between users is challenging \cite{ref90}. Thus, there should be much work to be done to realize an efficient UL MU-MIMO for the future WLANs.

\subsection{Group beamforming}

Group-based beam indication can reduce signalling/feedback overhead and allow certain flexibility of using beams within a group for transmission/reception \cite{ref91}, \cite{ref92}. Group-based beam switching can be supported when multiple beam groups are maintained, which means that beam recovery can be achieved via group switching instead of re-initiating initial access procedure in the case of blockage. To capture the relationship among different beams, new parameters, such as angle of arrival/departure, delay spread, average gain and Doppler shift spread, can be used to indicate the channel characteristics of beams in spatial domain. According to these parameters, one can group the beams into different subsets, where the beams in the same subset have similar parameters, and may share the same parameters which can significantly reduce the signalling or feedback overhead especially in multi-beam scenarios. How to come up with efficient beam grouping will be of great interest.

\subsection{Cloud-RAN based WLANs}

Resource management in cellular networks is a classical research topic. The centralized control of cellular networks makes it easy to carry out resource allocation. Nevertheless, it is difficult to manage wireless resource in WLANs, because all STAs operate in a distributed manner. Interference is also a major problem in WLANs that will limit the performance of WLANs. Thus, based on the aforementioned open research issues (e.g., inter-AP cooperation, dual connectivity, etc.) of Section VII, we believe that the future WLANs will probably be evolved to use Cloud-RAN based centralized control. More specifically, there will be a controller in the cloud, and a large number of APs are connected to a controller. In order to mitigate interference, those beams belong to different APs that interfere with each other can be allocated to operate in different time or different channels by the controller. In addition, based on the position information and moving direction of an STA, the controller can predict the location of available links and let the corresponding APs prepare the subsequent data delivery for the STA. Thus, the delay of AP handoff can be reduced. In the proposed Cloud-RAN based WLAN architecture, efficient and smart algorithms can be designed to address resource management, beam management, interference management, and mobility management.

\section{Conclusions}
Exploding mobile traffic due to emerging Internet of Things and smart city applications demands more spectrum resource and new technologies and mmWave communications seem to hold the key to deliver the solution. Therefore, mmWave communications have attracted intensive attention in both future 5G cellular systems and WLANs. There are several emerging WLAN standards that have been designed for mmWave frequency bands such as the IEEE 802.15.3c and IEEE 802.11ad. The IEEE 802.11ay is considered as an enhancement for the IEEE 802.11ad, attempting to provide ultra-high data rate services by utilizing tremendous bandwidth in mmWave. In this survey, we highlight some important technologies in the IEEE 802.11ad and present some technical challenges for mmWave communications in the IEEE 802.11ay standardization activities. More specifically, we have described related technical aspects for IEEE 802.11ay, especially the channel access for multiple channels and beamforming for both SU-MIMO and MU-MIMO that are not supported in the 802.11ad, and elaborated needed efficient beamforming training mechanisms proposed as the enhancements of beamforming training to the IEEE 802.11ad. We also identified open issues and future research directions for future mmWave WLANs before the IEEE 802.11ay standard is finalized. We expect that this paper offers the up-to-date information on the IEEE 802.11ay standardization activities.

As a final remark, intuitively mmWave technologies will be more effective in low mobility environments. How to take full advantage of the huge bandwidth in mmWave and high data rate services is of paramount importance and this cannot be done without good network-level cooperation as we have alluded in this paper. Recently, a flexible cognitive capability harvesting network architecture has been developed to more effectively utilize idle licensed bands via cognitive radio network technologies (CCHN) \cite{ref93, ref94}, where mmWave technologies have also been included as the way to transport data of potentially large volume from data sources to the premises of data utilization. Under the CCHN architecture, it is envisioned that 802.11ad and 802.11ay may become the technologies to use to form the backhaul for high data rate communication services and powerful edge computing system.

\section*{Appendix A}

\begin{center}
Summary of Main Acronyms.
\end{center}

\begin{footnotesize}
\begin{center}
\tablefirsthead{\hline\multicolumn{1}{|c|}{\textbf{Acronyms}}&\multicolumn{1}{|c|}{\textbf{Definition}}\\}
\tablelasttail{\hline}
\tablehead{\hline\multicolumn{2}{|c|}
{\small\sl Continued from previous page}\\\hline
\multicolumn{1}{|c|}{\textbf{Acronyms}}&\multicolumn{1}{|c|}{\textbf{Definition}}\\}
\tabletail{\hline\multicolumn{2}{|c|}{\small\sl Continued on next page}\\\hline}
\begin{supertabular}{!{\vrule width0.6pt}c!{\vrule width0.6pt}p{6.2cm}<{\centering}!{\vrule width0.6pt}}
\Xhline{0.6pt}
ABT & Analog Beam Tracking \\
\Xhline{0.6pt}
A-BFT & Association Beamforming Training\\
\Xhline{0.6pt}
ACI & Adjacent Channel Interference\\
\Xhline{0.6pt}
APCH & Alternative Primary Channel\\
\Xhline{0.6pt}
ATI & Announcement Transmission Interval\\
\Xhline{0.6pt}
AWV & Antenna Weight Vector\\
\Xhline{0.6pt}
BC & Beam Combining\\
\Xhline{0.6pt}
BF & Beamforming\\
\Xhline{0.6pt}
BFT & Beamforming Training\\
\Xhline{0.6pt}
BHI & Beacon Header Interval\\
\Xhline{0.6pt}
BI & Beacon Interval\\
\Xhline{0.6pt}
BRP & Beam Refinement Protocol\\
\Xhline{0.6pt}
BRPIFS & BRP Interframe Space\\
\Xhline{0.6pt}
BSS & Basic Service Set\\
\Xhline{0.6pt}
BTI & Beacon Transmission Interval\\
\Xhline{0.6pt}
CBAP & Contention-Based Access Period\\
\Xhline{0.6pt}
CCA & Clear Channel Assessment\\
\Xhline{0.6pt}
CSI & Channel State Information\\
\Xhline{0.6pt}
CSMA/CA & Carrier Sense Multiple Access with Collision Avoidance\\
\Xhline{0.6pt}
CTS & Clear-to-Send\\
\Xhline{0.6pt}
C/U & Control plane/Data plane\\
\Xhline{0.6pt}
DAC/ADC & Digital-to-Analog Converter/Analog-to-Digital Converter\\
\Xhline{0.6pt}
DBC & Digital Baseband Channel Tracking\\
\Xhline{0.6pt}
DL & MU-MIMO	Downlink Multi-User MIMO\\
\Xhline{0.6pt}
DMG & Directional Multi-Gigabit\\
\Xhline{0.6pt}
DTI & Data Transfer Interval\\
\Xhline{0.6pt}
EDCF & Enhanced Distributed Coordination Function\\
\Xhline{0.6pt}
EDMG & Enhanced DMG\\
\Xhline{0.6pt}
eNodeB & evolved NodeB\\
\Xhline{0.6pt}
FST & Fast Session Transfer\\
\Xhline{0.6pt}
Gbps & Gigabit Per Second\\
\Xhline{0.6pt}
HF & High Frequency\\
\Xhline{0.6pt}
ILA & Intelligent Listening During A-BFT\\
\Xhline{0.6pt}
I-SMBT & Initiator SU-MIMO BF Training\\
\Xhline{0.6pt}
ISS & Initiator Sector Sweep\\
\Xhline{0.6pt}
I-TXSS & Initiator Transmit Sector Sweep\\
\Xhline{0.6pt}
L-CEF & Non-EDMG Channel Estimation Field\\
\Xhline{0.6pt}
LF & Low Frequency\\
\Xhline{0.6pt}
L-Header & Non-EDMG Header Field\\
\Xhline{0.6pt}
LOS & Line-of-Sight\\
\Xhline{0.6pt}
L-STF & Non-EDMG Short Training Field\\
\Xhline{0.6pt}
MAC & Medium Access Control\\
\Xhline{0.6pt}
MCS & Modulation and Coding Scheme\\
\Xhline{0.6pt}
MID & Multiple Sector ID Detection\\
\Xhline{0.6pt}
MIMO & Multiple Input Multiple Output\\
\Xhline{0.6pt}
mmWave & Millimeter-Wave\\
\Xhline{0.6pt}
NAV & Network Allocation Vector\\
\Xhline{0.6pt}
NLOS & Non-LOS\\
\Xhline{0.6pt}
OBSS & Overlap BSS\\
\Xhline{0.6pt}
OFDM & Orthogonal Frequency Division Multiplexing\\
\Xhline{0.6pt}
P2P & Point-to-Point\\
\Xhline{0.6pt}
P2MP & Point-to-Multi-Point\\
\Xhline{0.6pt}
PAPR & Peak to Average Power Ratio\\
\Xhline{0.6pt}
PBSS & Personal Basic Service Set\\
\Xhline{0.6pt}
PCP/AP & PBSS Control Point/Access Point\\
\Xhline{0.6pt}
PER & Packet Error Rate\\
\Xhline{0.6pt}
PHY & Physical Layer\\
\Xhline{0.6pt}
PIFS & Point Coordination Function Interframe Space\\
\Xhline{0.6pt}
PLCP & Physical Layer Convergence Protocol\\
\Xhline{0.6pt}
PPDU & PLCP Data Unit\\
\Xhline{0.6pt}
QoE & Quality-of-Experience\\
\Xhline{0.6pt}
QoS & Quality-of-Service\\
\Xhline{0.6pt}
R-SMBT & Responder SU-MIMO BF Training\\
\Xhline{0.6pt}
RSS & Responder Sector Sweep\\
\Xhline{0.6pt}
RTS & Request-to-Send\\
\Xhline{0.6pt}
R-TXSS & Responder Transmit Sector Sweep\\
\Xhline{0.6pt}
RX & Receive\\
\Xhline{0.6pt}
SA-BFT & Separated A-BFT\\
\Xhline{0.6pt}
SBA-BFT & Secondary Backoff A-BFT\\
\Xhline{0.6pt}
SIFS & Short Interframe Space\\
\Xhline{0.6pt}
SINR & Signal-to-Interference-Plus-Noise-Ratio\\
\Xhline{0.6pt}
SISO & Single Input Single Output\\
\Xhline{0.6pt}
SLS & Sector Level Sweep\\
\Xhline{0.6pt}
SNR & Signal-Noise-Ratio\\
\Xhline{0.6pt}
SP & Scheduled Service Period\\
\Xhline{0.6pt}
SPSH & Spatial Sharing\\
\Xhline{0.6pt}
SSW frame & Sector Sweep frame\\
\Xhline{0.6pt}
SSW-Ack & Sector Sweep Ack\\
\Xhline{0.6pt}
SSW-FBCK & Sector Sweep Feedback\\
\Xhline{0.6pt}
STA & Station\\
\Xhline{0.6pt}
SU-MIMO & Single-User MIMO\\
\Xhline{0.6pt}
TDMA & Time Division Multiple Access\\
\Xhline{0.6pt}
TRN & Training Sequences Field\\
\Xhline{0.6pt}
TRN-R & Receive Training\\
\Xhline{0.6pt}
TRN-T & Transmit Training\\
\Xhline{0.6pt}
TX & Transmit\\
\Xhline{0.6pt}
TXOP & Transmission Opportunity\\
\Xhline{0.6pt}
UL & Uplink\\
\Xhline{0.6pt}
Wi-Fi & Wireless Fidelity\\
\Xhline{0.6pt}
WLAN & Wireless Local Area Network\\
\Xhline{0.6pt}
\end{supertabular}
\end{center}
\end{footnotesize}

\ifCLASSOPTIONcaptionsoff
  \newpage
\fi

\end{document}